\documentclass[hyper]{JHEP3} 

\usepackage{epsfig}

\def\npb#1{Nucl.~Phys.~{\bf B#1}}

\def\mt{{\ifmmode M^{eff}_T\else $M^{eff}_T$\fi}}

\def\e{\epsilon}

\def\ra{\rangle}

\def\e3{$\epsilon_3$}

\def\ch2{$\chi^2$}

\def\co#1{{\ifmmode{\cal O}_{#1}\else${\cal O}_{#1}$\fi}}

\newdimen\unit
\def\point#1 #2 #3{\vbox to0pt{\kern-#2\unit
  \hbox{\kern#1\unit#3}\vss}
 \nointerlineskip}

\newcommand{\be}{\begin{equation}}
\newcommand{\ee}{\end{equation}}
\newcommand{\bea}{\begin{eqnarray}}
\newcommand{\eea}{\end{eqnarray}}

\newcommand{\msbar}{\overline{\mbox{MS}}}

\newcount\hour
\newcount\minute
\newtoks\amorpm
\hour=\time\divide\hour by60
\minute=\time{\multiply\hour by60 \global\advance\minute by-
\hour}
\edef\standardtime{{\ifnum\hour<12 \global\amorpm={am}%
    \else\global\amorpm={pm}\advance\hour by-12 \fi
    \ifnum\hour=0 \hour=12 \fi
    \number\hour:\ifnum\minute<100\fi\number\minute\the\amorpm}}
\edef\militarytime{\number\hour:\ifnum\minute<100\fi\number\minute}
\def\bold#1{\setbox0=\hbox{$#1$}%
     \kern-.025em\copy0\kern-\wd0
     \kern.05em\copy0\kern-\wd0
     \kern-.025em\raise.0433em\box0 }

\newcommand{\newc}{\newcommand}
\newc\eg{{\it {e.g.}}}  \newc\etal{{\it {et al.}}} \newc\ie{{\it i.e.}}
\newc\etc{{\it {etc}}}

\newcommand\lsim{\mathrel{\rlap{\lower4pt\hbox{\hskip1pt$\sim$}}
    \raise1pt\hbox{$<$}}}
\newcommand\gsim{\mathrel{\rlap{\lower4pt\hbox{\hskip1pt$\sim$}}
    \raise1pt\hbox{$>$}}}

\newc{\mhalf}{m_{1/2}}      \newc{\mzero}{m_0}

\newc{\tanb}{\tan\beta}
\newc{\azero}{A_0}
\newc{\at}{A_t} \newc{\ab}{A_b} \newc{\atau}{A_\tau}

\newc{\bmu}{B\mu}           \newc{\sgn}{{\rm sgn}}
\newc{\mone}{M_1}           \newc{\mtwo}{M_2}

\newc{\charone}{\chi_1^\pm} \newc{\mcharone}{m_{\chi_1^\pm}}

\newc{\hl}{h}               \newc{\mhl}{m_{\hl}}
\newc{\hh}{H}               \newc{\mhh}{m_{\hh}}
\newc{\ha}{A}               \newc{\mha}{m_{\ha}}
\newc{\hc}{H^{\pm}}         \newc{\mhc}{m_{\hc}}

\newc{\qzero}{Q_0}          \newc{\qstop}{Q_{\widetilde t}}

\newc{\amu}{a_{\mu}}        \newc{\amususy}{a_{\mu}^{\rm SUSY}}
\newc{\amuexpt}{a_{\mu}^{\rm expt}}        \newc{\amusm}{a_{\mu}^{\rm SM}}
\newc{\deltaamususy}{\Delta a_{\mu}^{\rm SUSY}}
\newc\gmtwo{(g-2)_{\mu}} \newc\deltaamu{\Delta a_{\mu}}

\newc{\yt}{h_t} \newc{\yb}{h_b} \newc{\ytau}{h_{\tau}}
\newc{\mtpole}{m_t^{\rm pole}} \newc{\mbpole}{m_b^{\rm pole}}
\newc{\mtaupole}{m_{\tau}^{\rm pole}}

\newc{\mtmtsmmsbar}{m_t(m_t)^{\msbar}_{{\rm SM}}}
\newc{\mtmtsmdrbar}{m_t(m_t)^{\drbar}_{{\rm SM}}}
\newc{\mtmtmssmdrbar}{m_t(m_t)^{\drbar}_{{\rm SUSY}}}

\newc{\mbmbsmmsbar}{m_b(m_b)^{\msbar}_{{\rm SM}}}
\newc{\mbmzsmmsbar}{m_b(\mz)^{\msbar}_{{\rm SM}}}
\newc{\mbmzsmdrbar}{m_b(\mz)^{\drbar}_{{\rm SM}}}
\newc{\mbmzmssmdrbar}{m_b(\mz)^{\drbar}_{{\rm SUSY}}}
\newc{\mtaumzsmmsbar}{m_{\tau}(\mz)^{\msbar}_{{\rm SM}}}
\newc{\mtaumzsmdrbar}{m_{\tau}(\mz)^{\drbar}_{{\rm SM}}}
\newc{\mtaumzmssmdrbar}{m_{\tau}(\mz)^{\drbar}_{{\rm SUSY}}}

\newc{\mgut}{M_{\rm GUT}}

\newc{\mplanck}{M_{\rm P}}      \newc{\mpl}{M_{\rm Pl}}
\newc{\msusy}{M_{\rm SUSY}}      \newc{\ms}{M_{\rm S}}

\newc{\jxf}{J({\xf})}
\newc{\jxfexact}{J_{\rm exact}({\xf})}  \newc{\jxfexp}{J_{\rm exp}({\xf})}
\newc{\VEV}[1]{\langle #1 \rangle}

\newc{\xf}{x_f}
\newc\vrel{v_{\rm rel}}
\newcommand\mchi{m_{\chi}}              

\newc\sell{{\widetilde e}_L}      \newc\msell{m_{\sell}}
\newc\selr{{\widetilde e}_R}      \newc\mselr{m_{\selr}}

\newc\snue{{\widetilde \nu}_e}      \newc\msnue{m_{\snue}}
\newc\snutau{{\widetilde \nu}_\tau}      \newc\msnutau{m_{\snutau}}

\newc\supl{{\widetilde u}_L}      \newc\msupl{m_{\supl}}
\newc\supr{{\widetilde u}_R}      \newc\msupr{m_{\supr}}
\newc\sdl{{\widetilde d}_L}      \newc\msdl{m_{\sdl}}
\newc\sdr{{\widetilde d}_R}      \newc\msdr{m_{\sdr}}

\newcommand\mgluino{m_{\widetilde g}}

\newc\hpm{H^\pm} \newc\hp{H^+} \newc\hm{H^-}
\newc\sfermion{\tilde f}  \newc\msfermion{m_{\sfermion}}

\newc\second{{\rm sec}}

\newc\alphas{\alpha_s}

\newc\alphaem{\alpha_{em}}

\newc{\gstar}{g_\ast}           \newc{\gsstar}{g_{s\ast}}
\newc{\geff}{g_{\rm eff}}

\newcommand\mz{m_{Z}}

\newc{\sthw}{\sin\theta_W}              \newc{\cthw}{\cos\theta_W}
\newc{\bino}{\widetilde B}              \newc{\wino}{\widetilde W_30}
\newc{\higgsinob}{{\widetilde H}^0_b}   \newc{\higgsinot}{{\widetilde H}^0_t}

\newc{\abund}{\Omega h^2}
\newc{\abundchi}{\Omega_\chi h^2}
\newc{\abundcdm}{\Omega_{CDM} h^2}
\newc{\omegam}{\Omega_{M}}       \newc{\abundm}{\Omega_{M} h^2}
\newc{\omegab}{\Omega_{b}}       \newc{\abundb}{\Omega_{b} h^2}
\newc{\omegacdm}{\Omega_{CDM}}
\newc{\omegatot}{\Omega_{TOT}}

\newc{\rhocrit}{\rho_{crit}}
\newc{\rhochi}{\rho_{\chi}}

\newc\br{\mbox{BR}}

\newc{\beq}{\begin{equation}}
\newc{\eeq}{\end{equation}}

\newcommand\vs{{\it {vs.}}}

\newc\stoponetwo{{\widetilde t}_{1,2}}
\newc\sbotonetwo{{\widetilde b}_{1,2}}
\newc\stauonetwo{{\widetilde \tau}_{1,2}}

\newc\bsgamma{b\ra s \gamma }
\newc\brbsgamma{\br( B\rightarrow X_s \gamma )}

\newc{\sigsip}{\sigma^{SI}_{p}} \newc{\sigsin}{\sigma^{SI}_{n}}
\newc{\sigsdp}{\sigma^{SD}_{p}} \newc{\sigsdn}{\sigma^{SD}_{n}}
\newc{\sigsiA}{\sigma^{SI}_{A}}

\long\def\begincomment#1\endcomment{%
        \begingroup\sf\baselineskip12pt#1\endgroup}

\title{Dark Matter And {\boldmath $B_s\rightarrow \mu^+\ \mu^-$}\\
With Minimal {\boldmath $SO_{10}$} Soft SUSY Breaking II}

\author{Radovan Derm\' \i \v sek\\
        Davis Institute for High Energy Physics,\\
    University of California, Davis, CA 95616, USA\\
    E-mail: \email{dermisek@physics.ucdavis.edu}}

\author{Stuart Raby\\
        Department of Physics, The Ohio State University, \\
174 W. 18th Ave., Columbus, Ohio  43210, USA\\
        E-mail: \email{raby@pacific.mps.ohio-state.edu}}

\author{Leszek Roszkowski\\
        Department of Physics and Astronomy, University of Sheffield,\\
        Sheffield S3 7RH, England\\
        E-mail: \email{L.Roszkowski@sheffield.ac.uk}}

\author{Roberto Ruiz de Austri\\
        Departamento de F\'{\i}sica Te\'{o}rica C-XI
    and Instituto de F\'{\i}sica Te\'{o}rica C-XVI,\\
    Universidad Aut\'{o}noma de Madrid, Cantoblanco,
 28049 Madrid, Spain\\
        E-mail: \email{rruiz@delta.ft.uam.es}}

\abstract{
We update and extend to larger masses our previous
analysis of the MSSM with minimal $SO_{10}$ [MSO$_{10}$SM] soft SUSY
breaking boundary conditions. We find a well--defined, narrow region
of parameter space which provides the observed relic density of dark
matter, in a domain selected to fit precision electroweak data,
including top, bottom and tau masses.  The model is highly constrained
which allows us to make several predictions. We find the light
Higgs mass $m_h \leq 121 \pm 3$ GeV and also upper bounds on the mass
of the gluino
$\mgluino\lsim3.1$~TeV and lightest neutralino $\mchi\lsim450$~GeV.
As the CP odd Higgs mass $m_A$ increases,
the region of parameter space consistent with WMAP data is forced to
larger values of $M_{1/2}$ and smaller values of $m_h$.  Hence, we
find an upper bound $m_A \lsim 1.3$ TeV. This in turn leads to
lower bounds on ${\rm BR}(B_s\rightarrow \mu^+\ \mu^-) > 10^{-8}$
(assuming minimal flavor violation) and
on the dark matter spin independent detection cross section $\sigsip > 10^{-9}$
pb. Finally, we extend our previous analysis to include WIMP signals
in indirect detection and find prospects for WIMP detection generally
much less promising than in direct WIMP searches.}


\preprint{OHSTPY-HEP-T-05-001\\UCD-05-10 }
\begin{document}


\section{ Introduction}\label{intro:sec}

The constrained minimal supersymmetric standard model [CMSSM]~\cite{cmssm} is a well defined model for soft SUSY
breaking with five independent parameters given by $m_0, \ M_{1/2}, \ A_0, \ $ $ \tan\beta$ and $sign(\mu)$.  It
has been used extensively for benchmark points for collider searches,
as well as for astrophysical and dark matter analyses.
The economy of parameters in this scheme makes it a useful tool for
exploring SUSY phenomena. However the CMSSM misses regions of soft
SUSY breaking parameter space which give qualitatively different
predictions. In a previous paper~\cite{Dermisek:2003vn} we considered
an alternate scheme, the minimal $SO_{10}$ supersymmetric model
[MSO$_{10}$SM]~\cite{bdr}, which is well motivated and opens up a
qualitatively new region of parameter space.
In light of the recent WMAP analysis and recent improved limits
from DZero and CDF on the branching
ratio ${\rm BR} (B_s \rightarrow \mu^+ \ \mu^-)$, and from CDMS
on the dark matter direct detection cross section,
we have decided to reanalyze the MSO$_{10}$SM,
including updated data and extending the region of
parameter space to larger values of $m_{16} \geq 3$ TeV
and $m_A \geq 500$ GeV.

The recent WMAP data~\cite{wmap} provides an important constraint on the model. The dark matter candidate in this
model is the lightest neutralino. However, since the scalar masses of the first two families are of order $m_{16}
> 1.2 \; {\rm TeV}$, and the third generation sfermions (except for the stops) also tend to be heavy, the usually
dominant annihilation channels, for the neutralino LSP to light fermions via $t$--channel sfermion exchange, are
suppressed. On the other hand, the process $\chi\chi\rightarrow f\bar f$ via $s$--channel CP odd Higgs $A$
exchange becomes important. This is due to the enhanced $A$ coupling to down--type fermions, which is
proportional to $\tan\beta$, and because, in contrast to heavy scalar exchange, the process is not $p$--wave
suppressed.

We also compute the branching ratio for the process
$B_s \rightarrow \mu^+ \ \mu^-$ due to $A$
exchange~\cite{bsmumu,Bobeth:2001sq}. It is absolutely essential to
include this latter constraint in our analysis,
particularly in light of the published DZero bound
${\rm BR} (B_s \rightarrow \mu^+ \ \mu^-) < 5.0 \times 10^{-7}
(4.1 \times 10^{-7})$ at 95\% (90\%) CL~\cite{Abazov:2004dj}
and the new preliminary CDF bound $< 2.0 \times
10^{-7}$ at 95\% CL~\cite{cdfbsmm05prelim} and DZero
bound $< 3.7 \times 10^{-7}$ at 95\%
CL~\cite{dzerobsmm05prelim}.

The CDMS Collaboration has recently improved the upper limit
on the spin independent dark matter WIMP elastic
scattering cross section on a proton, $\sigsip$, down to
$2\times 10^{-7}$~pb (at low WIMP mass)~\cite{cdms04limit}.
A further improvement by an order of magnitude is foreseen
within a year. We update our previous results for $\sigsip$
and show that a large fraction of the parameter space will be
probed with currently running detectors,
and will be completely explored with a new round of ``one--tonne''
detectors which plan to reach down to $\sigsip\gsim10^{-10}$~pb.

In addition, we now evaluate prospects for indirect detection of WIMPs
in muon flux from WIMP annihilation in the Sun, a gamma ray flux from
the Galactic center, as well as antiproton and positron fluxes from
the Galactic halo. Here detection prospects are somewhat varied and,
especially in the case of the gamma ray flux, strongly depend on an
adopted model of the Galactic halo.

The paper is organized as follows. In section~\ref{sec:m10ssm} we
review the MSO$_{10}$SM, describe its virtues and outline the
analysis. Then in section~\ref{results:sec} for different values of
$m_{16}$ and $m_A$ we compute ${\rm BR} (B_s \rightarrow \mu^+ \
\mu^-)$, the cosmological dark matter density $\abundchi$, \etc.
The major results of the paper are found in section~\ref{expt:sec}.
We discuss the upper bound on the CP odd Higgs mass, $m_A$, and
the resulting lower bound on ${\rm BR} (B_s \rightarrow \mu^+ \ \mu^-)$.
CDF and DZero can potentially probe the entire allowed range.
We also consider the predictions for underground dark matter searches.
The entire expected range can be fully explored at the next level
of underground dark matter searches.
We also consider indirect dark matter searches.
Finally in Table 1 we give Higgs and SUSY spectra relevant for
collider searches for four representative points in SUSY parameter
space.  We summarize our results in section \ref{sec:predictions}.


\section{ Minimal ${\bf SO_{10}}$ SUSY Model -- MSO$_{10}$SM }\label{sec:m10ssm}

\subsection{Framework}
\label{sec:framework}

Since there are several distinct theories in the literature called the
minimal SO(10) SUSY model, let us briefly describe here the properties
of our minimal $SO_{10}$ SUSY model [MSO$_{10}$SM]~\cite{bdr}. Quarks
and leptons of one family reside in the $\bf 16$ dimensional
representation, while the two Higgs doublets of the MSSM reside in one
$\bf 10$ dimensional representation.  For the third generation we
assume the minimal Yukawa coupling term given by $ {\bf \lambda \ 16 \
10 \ 16 }. $ On the other hand, for the first two generations and for
their mixing with the third, we assume a hierarchical mass matrix
structure due to effective higher dimensional operators.  Hence the
third generation Yukawa couplings satisfy $\lambda_t = \lambda_b =
\lambda_\tau = \lambda_{\nu_\tau} = {\bf \lambda}$.

Soft SUSY breaking parameters are also consistent with $SO_{10}$
with (1) a universal gaugino mass $M_{1/2}$, (2)
a universal squark and slepton mass $m_{16}$,\footnote{$SO_{10}$
does not require all sfermions to have the same mass.
This however may be enforced by non--abelian family symmetries or
possibly by the SUSY breaking mechanism.} (3) a universal scalar
Higgs mass $m_{10}$, and (4) a universal A parameter $A_0$.
In addition we have the supersymmetric (soft SUSY breaking)
Higgs mass parameters $\mu$ ($B \mu$).  $B \mu$ may, as in the
CMSSM, be exchanged for $\tan\beta$. Note, not all of these parameters
are independent. Indeed, in order to fit the low energy electroweak data,
including the third generation fermion masses, it has been shown
that $A_0, \ m_{10}, \ m_{16}$ must satisfy the constraints~\cite{bdr}
\bea A_0 \approx - 2 \ m_{16}; & m_{10} \approx \sqrt{2} \ m_{16}  
\label{eq:constraint1}
\\
m_{16} > 1.2 \; {\rm TeV}; & \mu, \ M_{1/2} \ll m_{16} 
\label{eq:constraint2}
\eea
with
\be \tan\beta \approx 50.
\label{eq:tanbeta}
\ee
This result has been confirmed by several independent analyses~\cite{Tobe:2003bc,Auto:2003ys}.\footnote{Note,
different regions of parameter space consistent with Yukawa unification have also been discussed
in~\cite{Tobe:2003bc,Auto:2003ys,Balazs:2003mm}.}   Although the conditions (Eqns.~\ref{eq:constraint1},
\ref{eq:constraint2}) are not obvious, it is however easy to see that (Eqn.~(\ref{eq:tanbeta})) is simply a
consequence of third generation Yukawa unification, since $m_t(m_t)/m_b(m_t) \sim \tan\beta$.

One loop threshold corrections at the GUT scale lead to two significant parameters we treat as free parameters,
although they are calculable in any GUT. The first is a correction to gauge coupling unification given by  \be
\epsilon_3 \equiv \left[\alpha_3(M_G) - \tilde \alpha_G\right]/\tilde \alpha_G \ee where the GUT scale $M_G$ is
defined as the scale where $\alpha_1(M_G) = \alpha_2(M_G) \equiv \tilde \alpha_G$. The second is a Higgs
splitting mass parameter defined by \be \Delta m_H^2 \equiv (m_{H_d}^2 - m_{H_u}^2)/2 m_{10}^2 . \ee In order to
fit the low energy data we find $\epsilon_3 \approx - 4\%$ and $\Delta m_H^2 \approx 13 \%$~\cite{bdr}. The
largest corrections to $\epsilon_3$ come from the Higgs and $SO_{10}$ breaking sectors,  while the correction to
$\Delta m_H^2$ is predominantly due to the right--handed $\tau$ neutrino. Note, for $M_{\bar \nu_\tau} \sim  5.8
\times 10^{13}$ GeV, the necessary Higgs splitting is reduced to 7\%~\cite{Dermisek:2005ij}.

Finally, as a bonus, these same values of soft SUSY breaking
parameters, with $m_{16} \gg$ TeV, result in two very interesting
consequences.  Firstly, it ``naturally" produces an inverted scalar
mass hierarchy [ISMH]~\cite{scrunching}. With an ISMH squarks and
sleptons of the first two generations obtain mass of order $m_{16}$ at
$M_Z$. The stop, sbottom, and stau, on the other hand, have mass less
than (or of order) a TeV. An ISMH has two virtues. (1) It preserves
``naturalness" (for values of $m_{16}$ which are not too large), since
only the third generation squarks and sleptons couple strongly to the
Higgs. (2) It ameliorates the SUSY CP and flavor problems, since these
constraints on CP violating angles or flavor violating squark and
slepton masses are strongest for the first two generations, yet they
are suppressed as $1/m_{16}^{2}$.  For $m_{16} > $ a few TeV, these
constraints are weakened~\cite{masieroetal,or1+2,for1+2}.  Secondly,
Super--Kamiokande bounds on $\tau(p \rightarrow K^+ \bar \nu) \ > 2.3
\times 10^{33}$ yrs~\cite{superk} constrain the contribution of
dimension 5 baryon and lepton number violating operators. These are
however minimized with $\mu, \ M_{1/2} \ll m_{16}$~\cite{pdecay}.


\subsection{Phenomenological Analysis}\label{sec:analysis}

We use a top--down approach with a global \ch2 analysis~\cite{chi2}.
The input parameters are defined by boundary
conditions at the GUT scale. The 11 input parameters at $M_G$ are
given by: three gauge parameters $M_G, \;
\alpha_G(M_G),$ $\epsilon_3$; the Yukawa coupling $\lambda$,
and 7 parameters, including $\mu,\;$ and the 6 soft
SUSY breaking parameters $M_{1/2},\; A_0,\; \tan\beta$ (replacing $B \mu$),
$\; m_{16}^2, \; m_{10}^2, \; \Delta m_H^2$.
These are fit in a global $\chi^2$ analysis defined in terms of
physical low energy observables. Note we keep three parameters
$ ( m_{16}, \ \mu, \ M_{1/2} )$ fixed;
while minimizing $\chi^2$ with the remaining 8 parameters.
Below we plot $\chi^2$ contours as a function of $\mu, \ M_{1/2}$
for different values of $m_{16}$.
We use two (one)~loop renormalization group [RG] running for
dimensionless (dimensionful) parameters from $M_G$ to $M_Z$.
\footnote{Note, we have checked that switching to 2 loop RGEs
for dimensionful parameters can be compensated for by small
changes in the GUT scale parameters, without significant changes
in the low energy results.}
We require electroweak symmetry breaking using an improved
Higgs potential, including $m_t^4$ and $m_b^4$ corrections
in an effective 2-Higgs doublet model below
$M_{SUSY} = \sqrt{\frac{1}{2} (m_{\tilde t_1}^2 +
m_{\tilde t_2}^2)}$~\cite{carenaetal,cqw}.

The $\chi^2$ function includes 9 observables; 6 precision electroweak
data $\alpha_{EM},$ $G_\mu,$ $\alpha_s(M_Z),$
$M_Z, \; M_W,$ $\rho_{NEW}$ and the 3 fermion masses
$M_{top},\; m_b(m_b), \; M_\tau$.
In our analysis we fit the central values~\cite{pdg2000}:
$M_Z = 91.188$ GeV, $M_W = 80.419$ GeV,
$G_{\mu}\times 105 = 1.1664$ GeV$^{-2}$,
$\alpha_{EM}^{-1} = 137.04,$ $M_{\tau} = 1.7770$ GeV
with 0.1\% numerical uncertainties;
and the following with the experimental uncertainty
in parentheses: $\alpha_s(M_Z) = 0.1172 \; (0.0020),$
$\rho_{new}\times 103 = -0.200\; (1.1)$~\cite{rhonew},
$M_t = 178.0 \; (4.3)$ GeV, $m_b(m_b) = 4.20\; (0.20)$ GeV.
\footnote{Note we take a conservative error for
$m_b(m_b)$~\cite{pdg2000} in view of recent claims to much
smaller error bars~\cite{Beneke:1999fe}.}
We include the complete one loop threshold corrections at $M_Z$ to all
observables. In addition we use one loop QED and three
loop QCD RG running below $M_Z$.

The output of this analysis is a set of weak scale squark, slepton,
gaugino and Higgs masses.  With regards to
the calculated Higgs and sparticle masses, the neutral Higgs masses
$h,\; H, \; A$ are pole masses calculated
with the leading top, bottom, stop, sbottom loop contributions;
while all other sparticle masses are running masses.
This output is then used to compute the cosmological dark matter density of
the lightest neutralino $\abundchi$, which is the LSP,
the branching ratio ${\rm BR} (B_s \rightarrow \mu^+ \ \mu^-)$, and other
observables, as described in more detail below.
Note, it is important to emphasize that neither $\abundchi$ nor
the branching ratio ${\rm BR} (B_s \rightarrow \mu^+ \ \mu^-)$
are included in the $\chi^2$ analysis.
They are predictions of the model obtained in the regions
selected by consistency with gauge coupling and third generation
Yukawa unification and the low energy observables.

Using $\chi^2$ penalties\footnote{In order to constrain the values
of some physical observables in our $\chi^2$ analysis, such as
$m_{\tilde t_1}$ or $m_{A}$, we add a significant contribution to
the $\chi^2$ function for values of these observables outside the
desired range. We refer to this additional contribution as a
$\chi^2$ penalty.   Minimization of $\chi^2$ with Minuit, then
pushes the fits to the desired range.   Of course the $\chi^2$
penalties then vanish.} we apply two additional constraints:

\begin{itemize}
\item  $m_{\tilde t_1} \geq 300$  GeV
\item  $m_{A}$ fixed.
\end{itemize}

The first is chosen to be consistent with ${\rm BR} (B\rightarrow
X_s\gamma)$~\cite{bdr}.  Note, although we do calculate ${\rm BR}
(B\rightarrow X_s\gamma)$, we do not use it as a constraint in the
analysis. This is for two reasons --- 1) this decay mode depends
on 3--2 generation mixing which is model dependent and 2) it is
not difficult to fit ${\rm BR} (B\rightarrow X_s\gamma)$ for
large enough values of $m_{\tilde t_1}$.
Hence, in order to be
generally consistent with the measured value of ${\rm BR}
(B\rightarrow X_s\gamma)$, we impose $m_{\tilde t_1} \geq 300$
GeV.  With regards to the second constraint, since $\abundchi$ and
${\rm BR} (B_s \rightarrow \mu^+ \ \mu^-)$ are both sensitive to
the value of $m_{A}$, we fix its value and present our results
for different values of $m_{A}$.\footnote{The calculation of ${\rm
BR} (B\rightarrow X_s\gamma)$ and ${\rm BR} (B_s \rightarrow \mu^+
\ \mu^-)$ requires a model for fermion mass matrices.  In the
absence of such a model we use the observed CKM matrix elements to
calculate these flavor violating branching ratios.}

Finally, after performing the $\chi^2$ analysis, we impose the following constraints on the parameter space:
\begin{itemize}
\item
lower bound on the
lightest chargino mass $m_{\chi^+} > 104$~GeV.

\item
lower bound on the light Higgs mass  $m_h > 111$~GeV.
Note, because of the theoretical uncertainty in the calculation of
$m_h$ ($\sim3$~GeV), we conservatively impose $m_h>111$~GeV,
instead of the LEP bound for SM Higgs $m_h>114.4$~GeV.

\item branching ratio ${\rm BR}(B_s\rightarrow \mu^+\ \mu^-)$. The current best published limit comes from DZero
${\rm BR} (B_s \rightarrow \mu^+ \ \mu^-) < 5.0 \times 10^{-7} (4.1 \times 10^{-7})$ at 95\% (90\%)
CL~\cite{Abazov:2004dj}. Recently CDF has announced a new preliminary bound $< 2.0 \times 10^{-7}$ at 95\%
CL~\cite{cdfbsmm05prelim}, while DZero has come up with a new preliminary bound $< 3.7 \times 10^{-7}$ at 95\%
CL~\cite{dzerobsmm05prelim}. We'll discuss the impact of the bounds below.

\item the relic abundance of the lightest neutralino in the range $0.094< \abundchi < 0.129$ (2~$\sigma$) which
we take as a preferred range. We will exclude points for which $\abundchi > 0.129$ but allow for the possibility
that $\abundchi < 0.094$ (subdominant component of cold DM).
\end{itemize}

The quantity ${\rm BR}(B_s\rightarrow \mu^+\ \mu^-)$ is computed assuming the CKM mixings among squarks (minimal
flavor violation). For this purpose we use leading log expressions derived in the first paper of
Ref.~\cite{Bobeth:2001sq}. The SM prediction is around $3\times 10^{-9}$ while SUSY contributions to the process
are dominated by pseudoscalar exchange and scale as $\tan^6\beta/m^4_A$, and can be large. Note, the branching
ratio also depends on the (model dependent) 3--2 generation squark mixings. However, the mixings tend to increase
${\rm BR}(B_s\rightarrow \mu^+\ \mu^-)$ which make our lower bound on $m_A$ stronger. On the other hand, beyond
leading log corrections can relax our bounds on $m_A$ by up to ${\cal O} (20\%)$ because of a ``focusing
effect''~\cite{or1+2,for1+2}, especially if the squark mixings are non--minimal.

We compute the relic abundance $\abundchi$ of the lightest neutralino
using exact expressions for neutralino pair annihilation into all
allowed final--state channels, which are valid both near and further
away from resonances and thresholds~\cite{nrr1+2}. We further treat
the neutralino coannihilation with the lightest chargino and
next--to--lightest neutralino~\cite{eg97} and with the lighter
stau~\cite{nrr3} with similar precision.  We include all
coannihilation channels, including the previously
neglected~\cite{Dermisek:2003vn} coannihilation with light stops,
although this channel only affects points in the parameter space which
are excluded by other constraints. We solve the
Boltzmann equation numerically as in~\cite{darksusy} and compute
$\abundchi$ with an error of a few per cent, which is comparable with
today's accuracy on the observational side.

\begin{figure}[t!]
\begin{center}
\begin{minipage}{6in}
\epsfig{file=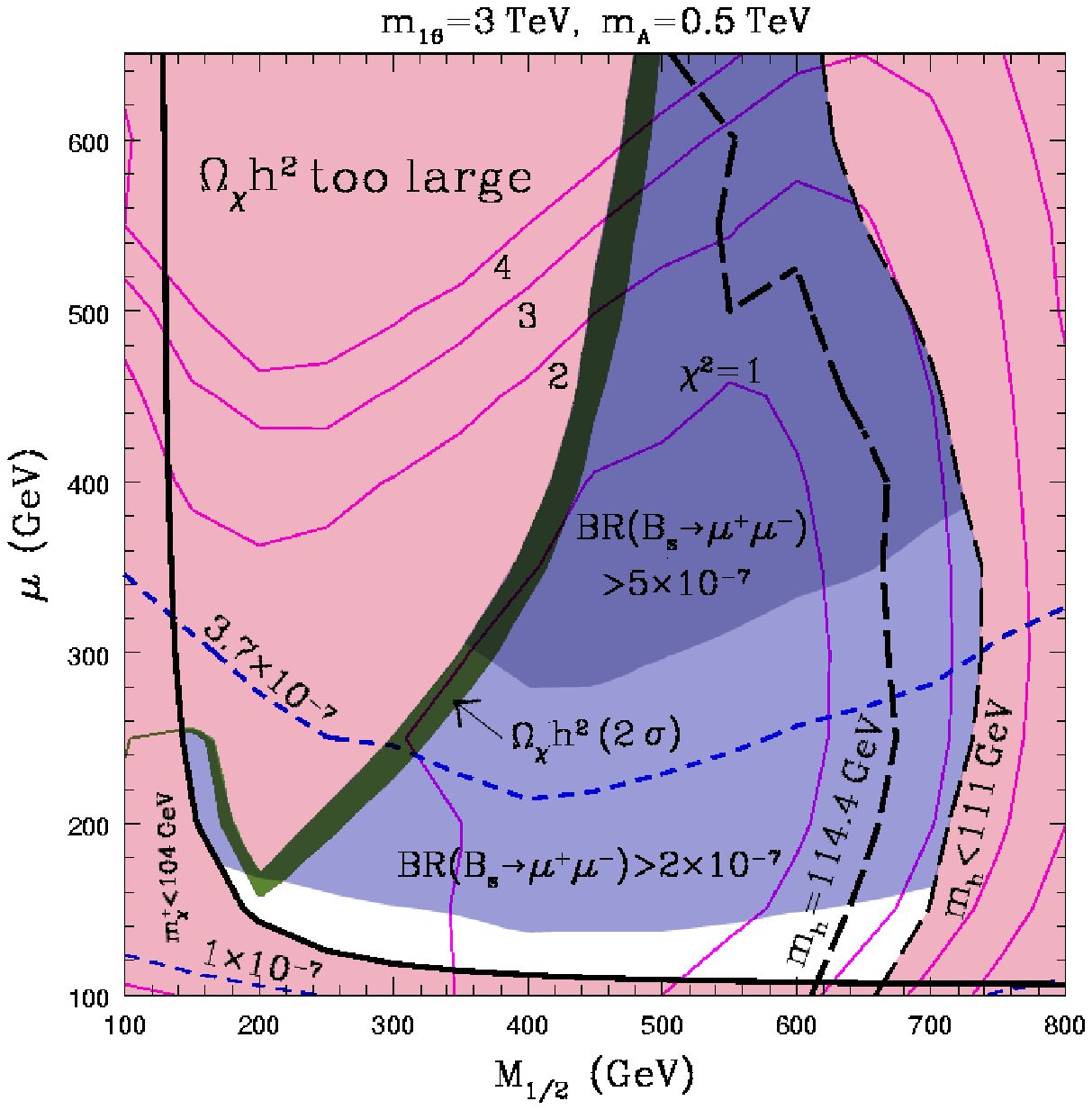,width=3in} \hspace*{-0.15in}
\epsfig{file=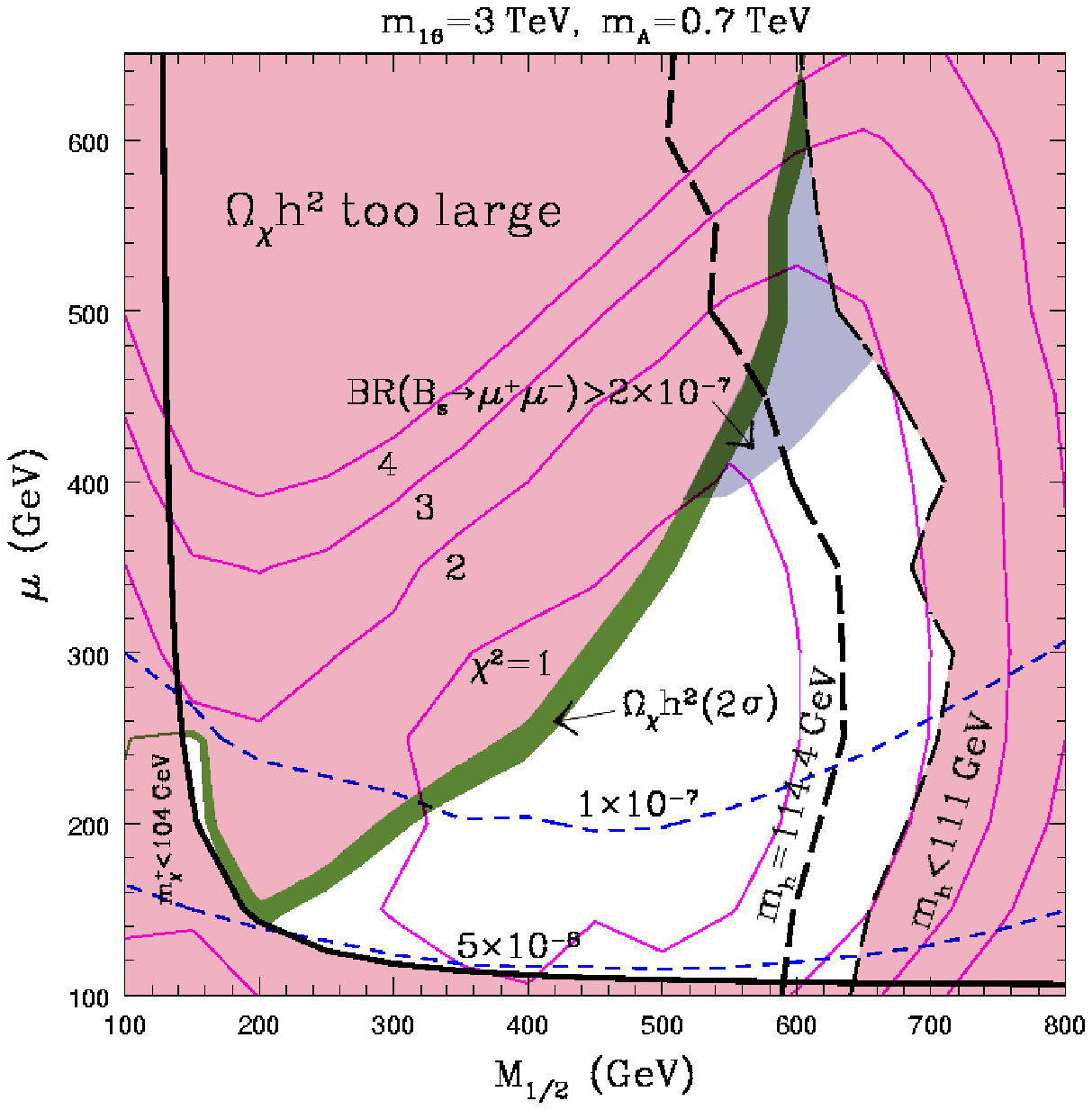,width=3in}
\end{minipage}
\end{center}
\begin{center}
\caption{\label{fig:3kall} {\small Contours of constant $\chi^2$ for
    $m_{16} = 3$ TeV and $m_{A} = 500$ GeV (left panel) and 700 GeV (right
    panel). The red (light shaded) regions are excluded by $m_{\chi^+}<104$ GeV
    (below and to the left of the solid black curve), $m_{h}<111$ GeV
    (on the right) and by $\abundchi>0.129$. To the right of the thick
    broken black line one has $m_{h}<114.4$ GeV. The green (darkest
    shaded) band
    corresponds to the preferred $2\,\sigma$ range $0.094< \abundchi <
    0.129$, while the white regions below it correspond to
    $\abundchi<0.094$.  The region excluded by the DZero experimental
    bound on ${\rm BR} (B_s \rightarrow \mu^+ \ \mu^-)
    < 5.0 \times 10^{-7}$ at 95\% CL is marked
    in dark blue (dark shaded), while the region affected by
    the new preliminary CDF bound $< 2.0 \times 10^{-7}$ at 95\% is marked
    in light blue (light shaded).  Contours of constant ${\rm BR}
    (B_s \rightarrow \mu^+ \ \mu^-)$ are given by the blue dashed  lines.} }
\end{center}
\end{figure}


\section{Results}\label{results:sec}

\begin{figure}[t!]
\begin{center}
\begin{minipage}{6in}
\epsfig{file=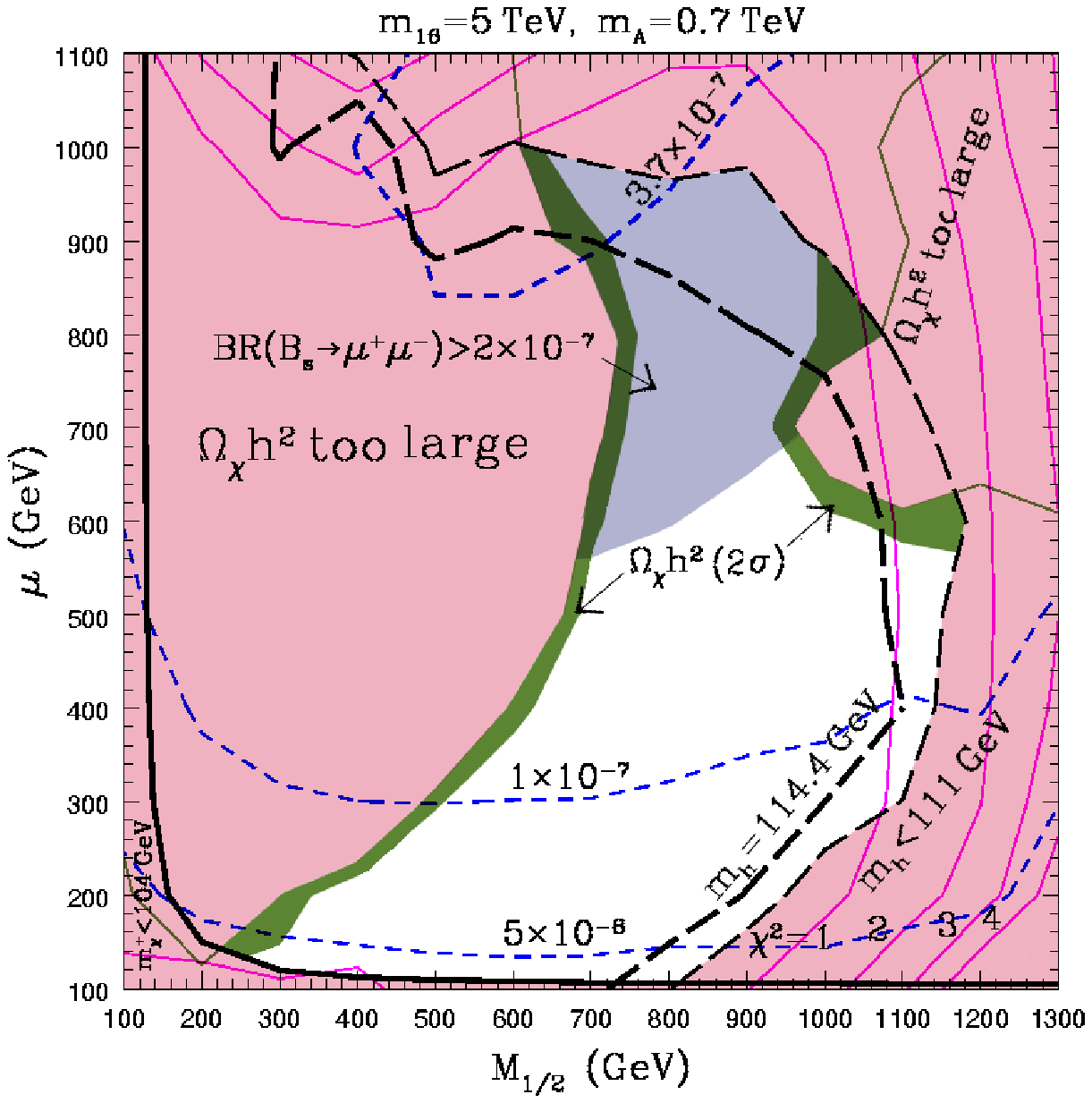,width=3in} \hspace*{-0.15in}
\epsfig{file=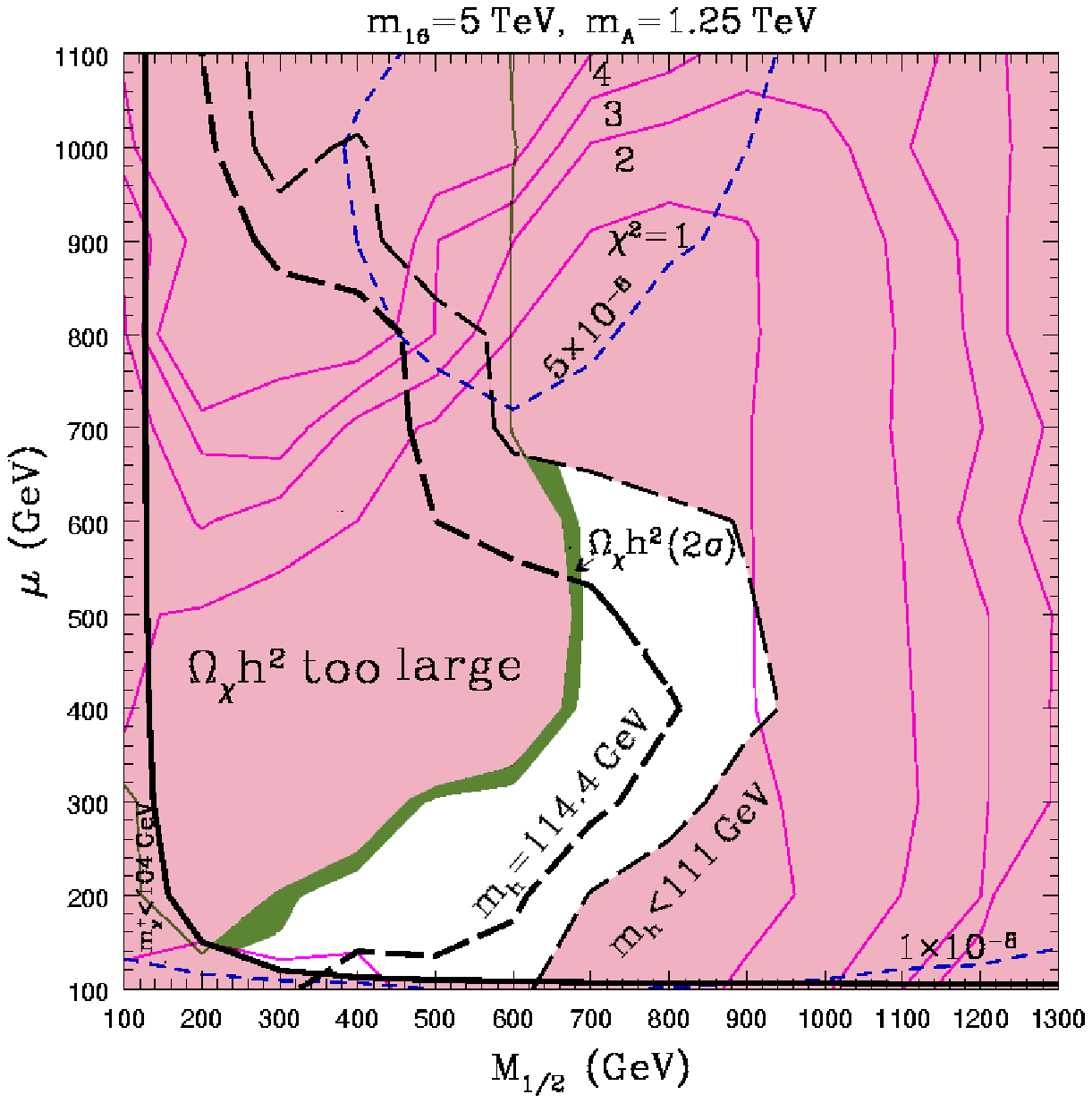,width=3in}
\end{minipage}
\caption{\label{fig:5kall} {\small Same as
    Fig.~{\protect\ref{fig:3kall}}  for $m_{16} =
5$ TeV and $m_{A} = 700$~GeV (left window)  and 1250~GeV (right window).} }
\end{center}
\end{figure}

\begin{figure}[t!]
\begin{center}
\begin{minipage}{6in}
\epsfig{file=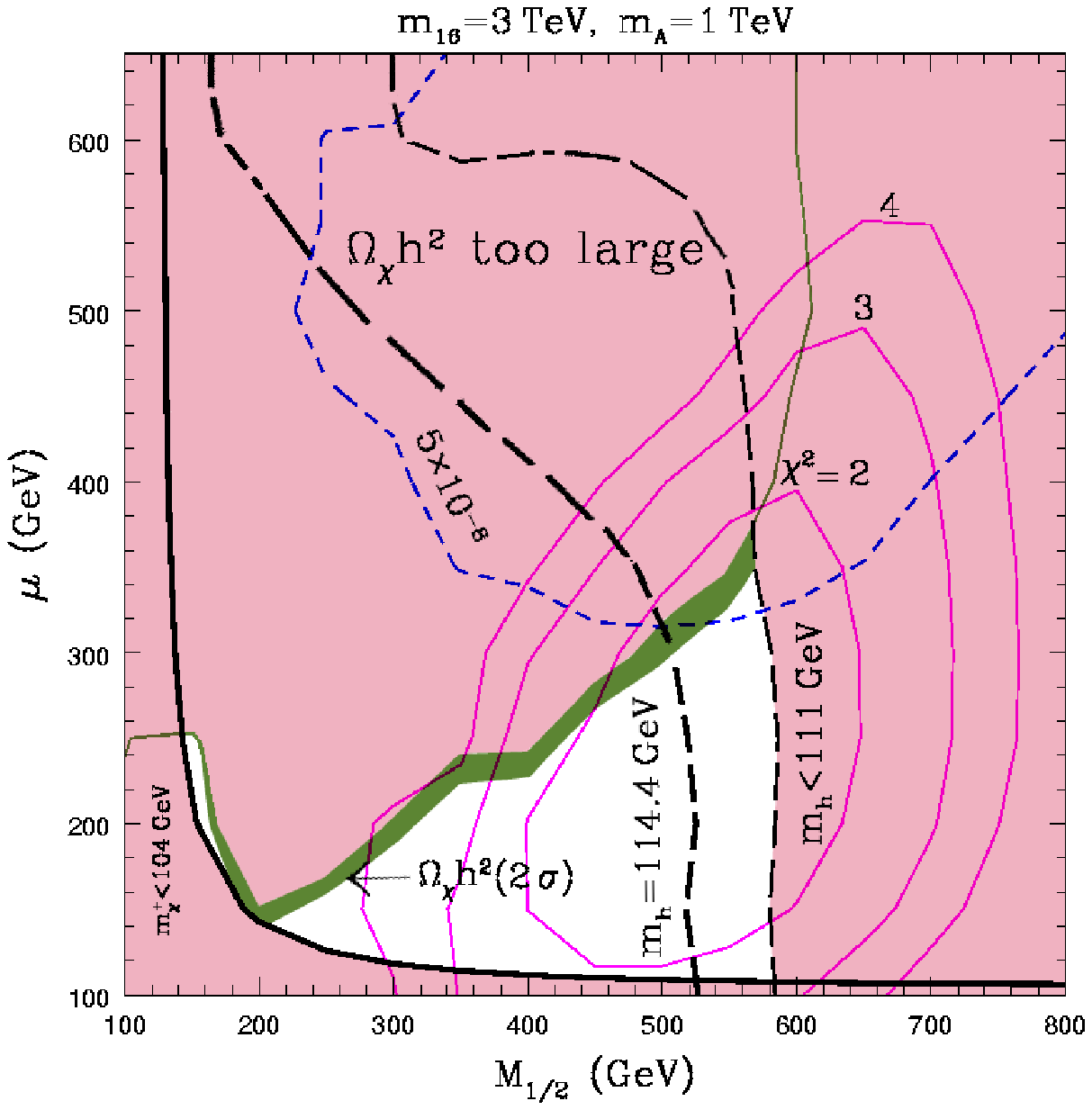,width=3in} \hspace*{-0.15in}
\epsfig{file=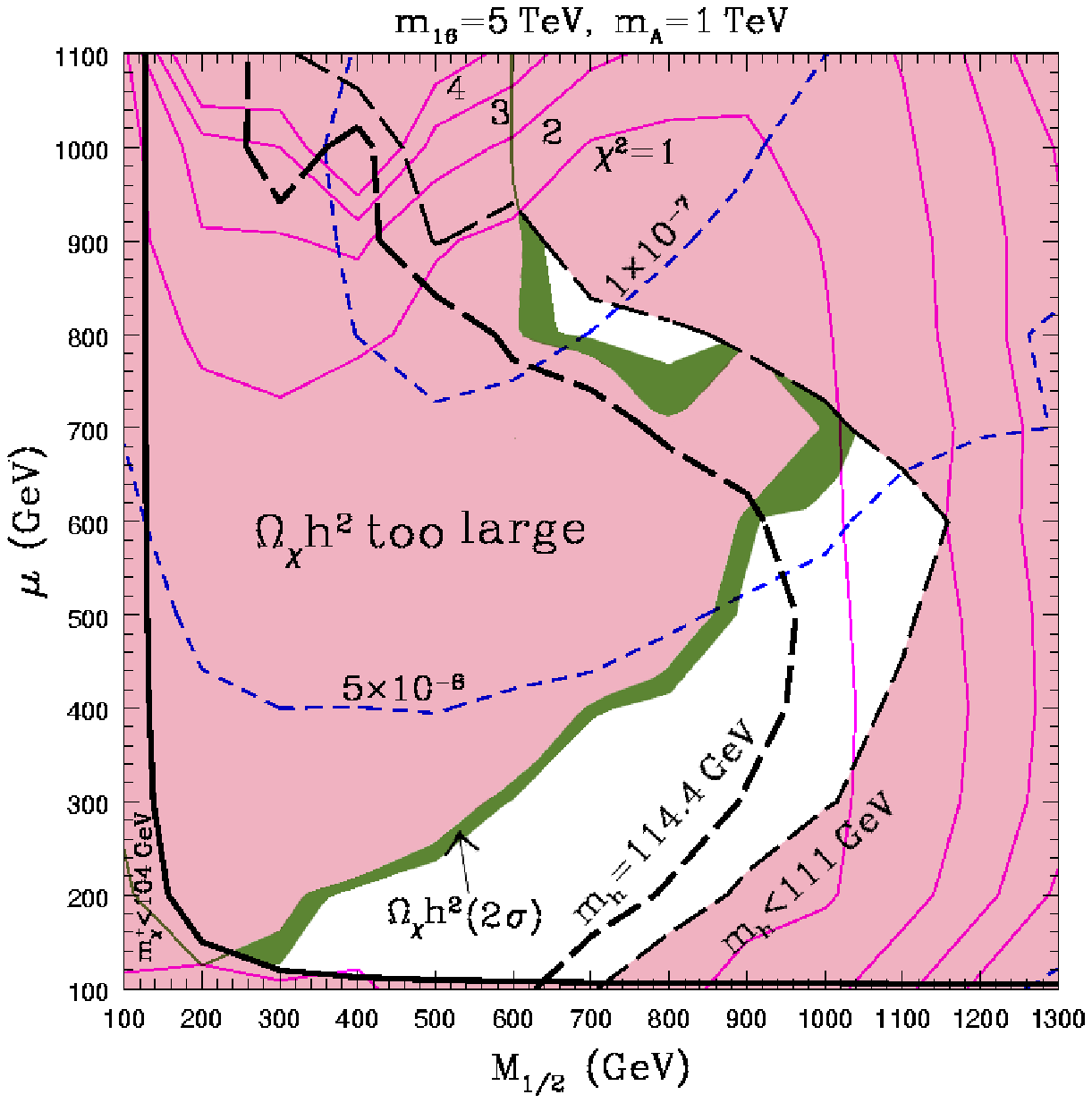,width=3in}
\end{minipage}
\caption{\label{fig:3k5k1k} {\small Same as
    Fig.~{\protect\ref{fig:3kall}}  for $m_{16} =
3$ TeV (left window) 5~TeV (right window) and $m_{A} = 1$ TeV.} }
\end{center}
\end{figure}

In Figs.~\ref{fig:3kall},~\ref{fig:5kall} and \ref{fig:3k5k1k} we present our results for different values of
$m_{16}$ and $m_{A}$ in the $\mu, \ M_{1/2}$ plane.  For example, in Fig.~\ref{fig:3kall} we present, for $m_{16}
= 3$ TeV and $m_{A} = 500$ GeV (left panel) and $m_{A} = 700$ GeV (right panel) the (magenta) solid lines of
constant $\chi^2$. The red (lightest shaded) regions are excluded by collider limits and by $\abundchi > 0.129$.
Specifically the red regions at small $\mu$ and/or small $M_{1/2}$ (bounded by the solid black line) is excluded
by bounds on the chargino mass $m_\chi > 104$ GeV and the red region on the right side (bounded by the dashed
black line) is excluded by (our conservative application of) the LEP Higgs bound $m_h > 111$ GeV. The
cosmologically preferred dark matter region green (darkest shaded) satisfy $0.094 < \abundchi < 0.129$. We find
significant regions of parameter space which give $\chi^2 \leq 3$, satisfy the preferred $\abundchi$ range as
above, and satisfy all other phenomenological constraints. In Fig.~\ref{fig:5kall}, we have $m_{16} = 5$ TeV and
$m_{A} = 700$ and 1250~GeV, respectively, and in Fig.~\ref{fig:3k5k1k} we have $m_{16} = 3$ and 5 TeV,
respectively, and $m_{A} = 1$~TeV.

In Fig.~\ref{fig:3kall}, marked in dark blue (dark shaded), is the region excluded by the DZero experimental
bound on ${\rm BR} (B_s \rightarrow \mu^+ \ \mu^-) < 5.0 \times 10^{-7}$ at 95\% CL~\cite{Abazov:2004dj} and, in
Figs.~\ref{fig:3kall} and \ref{fig:5kall}, light blue (light shaded) is the region affected by the new
preliminary CDF bound $< 2.0 \times 10^{-7}$ at 95\% CL~\cite{cdfbsmm05prelim}. In addition, we have included
contours of constant ${\rm BR} (B_s \rightarrow \mu^+ \ \mu^-)$ with dashed blue lines. The branching ratio ${\rm
BR} (B_s \rightarrow \mu^+ \ \mu^-)$ is sensitive to the value of the CP odd Higgs mass $m_{A}$~\cite{bsmumu},
scaling as $m_A^{-4}$.  For $m_{A} = 500$ GeV (Fig.~\ref{fig:3kall} (left)) the branching ratio ${\rm BR} (B_s
\rightarrow \mu^+ \ \mu^-)$ is below the published DZero bound, for acceptable values of $\abundchi$ and $\chi^2
< 3$, but is almost excluded by the preliminary CDF bound.  While for $m_A = 1.25$ TeV (see Fig.~\ref{fig:5kall}
(right window)) we have ${\rm BR} (B_s \rightarrow \mu^+ \ \mu^-) > 10^{-8}$.

The cosmological relic abundance of the neutralino $\abundchi$
is primarily determined by the direct s--channel
pair--annihilation into SM fermion pairs through the CP odd Higgs.
Since all the sfermions are very heavy, their contribution
to reducing the neutralino number density is strongly suppressed.
In contrast, because of the coupling $A b{\bar b}\propto \tan\beta$
(and similarly for the $\tau$'s), the $A$--resonance is effective and
broad. Near $\mchi\approx m_{A}/2$ it reduces $\abundchi$ down to allowed
but uninterestingly small values $\ll 0.1$.
As one moves away from the resonance, $\abundchi$ grows,
reaches the preferred range $0.094 < \abundchi < 0.129$, before becoming
too large $\abundchi > 0.129$.
\footnote{Note that at one loop we have $M_1(M_Z) = M_{1/2}
* \alpha_1(M_Z)/\alpha_G$ so $M_1(M_Z) \approx 0.4 M_{1/2}$.
For bino--like neutralino (which is true for larger $\mu$),
we thus have $\mchi \approx 0.4 M_{1/2}$.
Hence for s--channel annihilation we have $m_{A} \approx 2
m_{\chi} \approx 0.8 \ M_{1/2}$ or $M_{1/2}\approx (5/4)\,m_{A}$
for the position of the ``peak suppression."}
(A similar, but much more narrow resonance due to $h^0$ is also present at
$M_{1/2}\approx 150$~GeV and small $\mu$.)
When $\mchi\gsim m_t$ ($M_{1/2}\gsim420$~GeV) and the stops
are not too heavy, the LSP pairs annihilate to $t\,{\bar t}$--pairs.
In the region of large $M_{1/2}$, often where $m_h$ is already too low,
two additional channels become effective.
First, in this region the neutralino becomes almost mass degenerate
with the lighter stau which leads to reducing $\abundchi$
through coannihilation. Second, if $m_{A}$ is not too large, neutralino
pair--annihilation into Higgs boson pairs $AA$ and $HH$ opens up.
Finally, at $\mu\ll M_{1/2}$, the relic abundance is strongly reduced
due to the increasing higgsino component of the LSP.

\begin{figure}[t!]
\begin{center}
\begin{minipage}{5in}
\epsfig{file=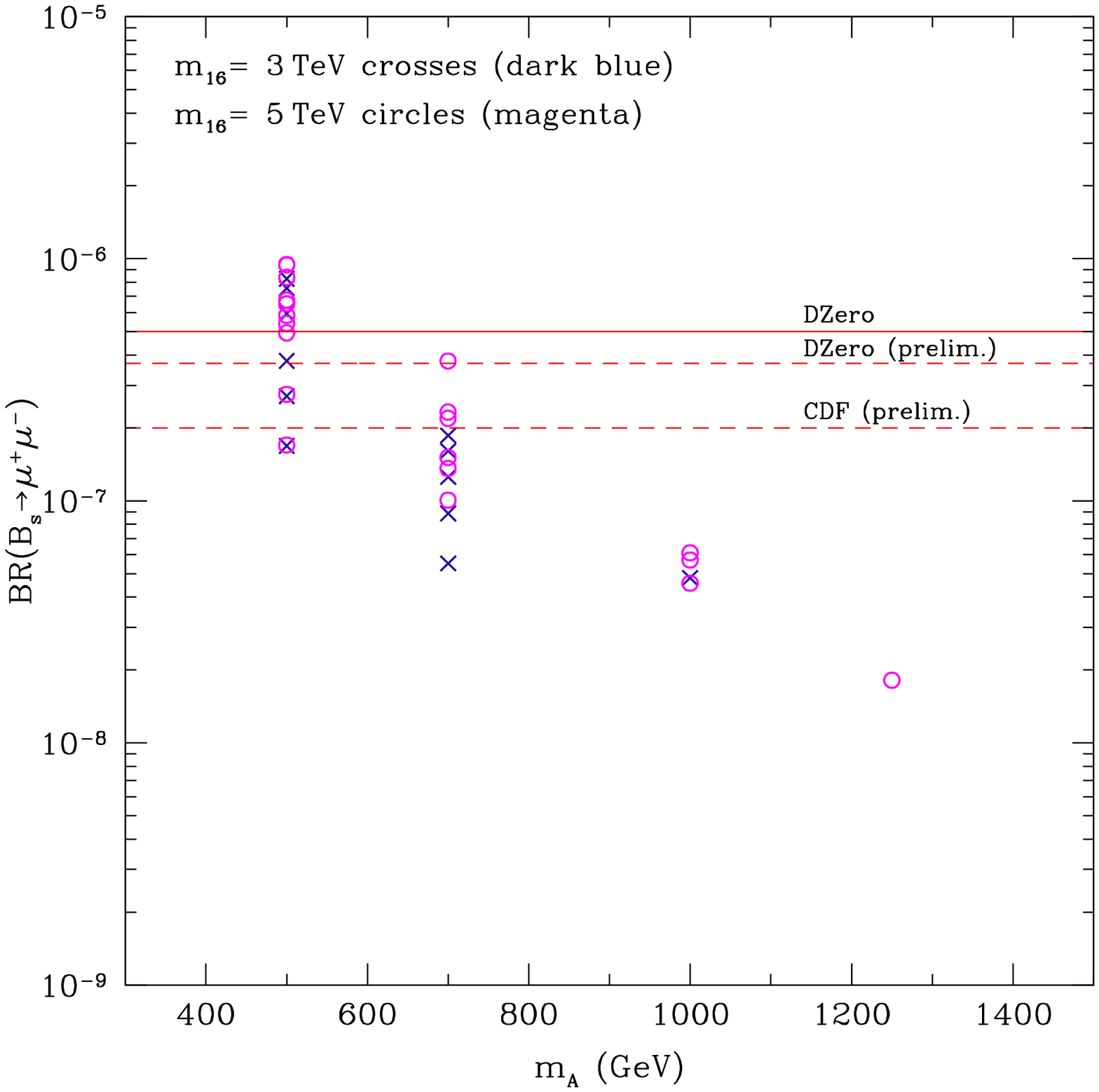,height=5in}
\end{minipage}
\caption{\label{fig:bsmm-ma} {\small The branching ratio
${\rm BR} (B_s \rightarrow \mu^+\ \mu^-)$ as a function
of $m_A$ in the $\mu, \ M_{1/2}$ region of parameter space
satisfying all the collider constraints, $0.094 <
\abundchi < 0.129$ and $\chi^2 < 3$ for fixed $m_{16}$ = 3 and 5~TeV.}}
\end{center}
\end{figure}

\subsection{Experimental Tests}\label{expt:sec}

\noindent{\boldmath $\underline{BR(B_s \rightarrow \mu^+ \mu^-)}$.} \hspace*{0.1in} As we see in
Figs.~\ref{fig:3kall}--\ref{fig:5kall}, the preliminary Tevatron limit on $BR(B_s \rightarrow \mu^+ \mu^-)$ puts
a significant lower bound on $m_A \gsim 500$~GeV. The dependence on $m_A$ is plotted in Fig.~\ref{fig:bsmm-ma}
with points in the preferred region with all the collider constraints satisfied, $\chi^2 < 3$ and $0.094
<\abundchi <0.129$, for the two representative cases $m_{16} = 3$ and 5 TeV. Note that, for a fixed value of
$m_A$, a vertical spread of points is caused by the changing shape of the cosmologically favored band in the
$\mu, \ M_{1/2}$ plane and some variation in $\tan\beta$ in Figs.~\ref{fig:3kall}--\ref{fig:3k5k1k}.

We now argue that there is an upper bound on $m_A$, the CP odd Higgs mass. Hence there is a lower bound on the
branching ratio $BR(B_s \rightarrow \mu^+ \mu^-)$. We first show that increasing the value of $m_{16}$ permits a
larger range for the parameters $\mu, \; M_{1/2}$.  Comparing Figs.~\ref{fig:3kall}--\ref{fig:3k5k1k} we see
that, as $m_{16}$ increases, the region with $\chi^2 < 3$ rapidly grows. Note, the dominant pull in $\chi^2$ is
due to the bottom quark mass. In order to fit the data, the total SUSY corrections to $m_b(m_b)$ must be of order
$-(2 - 4) \%$~\cite{bdr}. In addition there are three dominant contributions to these SUSY corrections: a gluino
loop contribution $\propto \ \alpha_3 \ \mu \ M_{\tilde g} \ \tan\beta/m_{\tilde b_1}^2$, a chargino loop
contribution $\propto \ \lambda_t^2 \ \mu \ A_t \ \tan\beta/m_{\tilde t_1}^2$, and a term $\propto \ \log
M_{SUSY}^2$.  Larger values of $m_{16}$ permit a larger range for the ratio $m_{\tilde b_1}/m_{\tilde t_1}$. {\em
Thus larger values of $m_{16}$ allows more freedom in parameter space for fitting the data at both smaller or
larger values of $\mu, \; M_{1/2}$.}  Note, also when $m_{16}$ increases (with $M_{1/2}$ fixed) the parameter
$A_t$ becomes more negative, since $A_0 \approx - 2 \ m_{16}$ and $A_t \approx - 3 \ M_{1/2} + \epsilon \ A_0$
where $\epsilon \ll 1$.

In addition, as $M_{1/2}$ increases the light Higgs mass decreases. The reason for this can easily be seen.
Consider the approximate equation for the light Higgs mass (Eqn.~(2.32) from Ref.~\cite{cqw})~\footnote{This
approximation gives a value for the light Higgs mass which is larger than the value obtained with a more exact
calculation (see Fig. 2 in Ref.~\cite{cqw}) for values of $|A_t|$ above 2 TeV. The discrepancy increases with
increasing $|A_t| \geq 2$ TeV, overestimating the Higgs mass by as much as 8 GeV for $|A_t| = 3.6$ TeV.} \bea
m_h^2 \approx & M_Z^2 \left( 1 - \frac{3}{8 \pi^2} \frac{m_t(m_t)^2}{v^2} t \right) & \nonumber \\ & + \frac{3}{4
\pi^2} \frac{m_t(m_t)^4}{v^2} \left[ \frac{1}{2} \tilde X_t + t + \frac{1}{16 \pi^2} \left( \frac{3}{2}
\frac{m_t(m_t)^2}{v^2} - 32 \pi \alpha_3 \right) (\tilde X_t t + t^2) \right] & \eea where $\tilde X_t = \frac{2
A_t^2}{M_{SUSY}^2} \left(1 - \frac{A_t^2}{12 M_{SUSY}^2} \right)$, $\;\; t = \log(M_{SUSY}^2/m_t(m_t)^2) \;\;$
and $\;\; M_{SUSY} \approx \sqrt{(m_{\tilde t_1}^2 + m_{\tilde t_2}^2)/2}$. In our analysis, the bottom quark
gets a positive supersymmetric correction proportional to $\mu M_{1/2}$ and a negative contribution proportional
to $\mu A_t$.  Thus in order to fit the bottom quark mass as $M_{1/2}$ increases, $- A_t$ also increases. As a
consequence, in the relevant region of parameter space, $\tilde X_t$ decreases. In addition, the fit value of
$m_t(m_t)$ also decreases. {\em  Hence the value of $m_h$ decreases as $M_{1/2}$ increases.}  On the other hand,
the value of the light Higgs mass is fairly insensitive to $m_{16}$.\footnote{Although for fixed $M_{1/2}$ the
light Higgs mass does increase by a small amount as $m_{16}$ increases (see Fig.~\ref{fig:3k5k1k}).}  In the
acceptable regions of parameter space we find $114.4 < m_h < 121$ GeV. \footnote{Note, there is at least a $\pm
3$ GeV theoretical uncertainty in the Higgs mass.}

Now let us put it all together. Comparing various windows in Figs. \ref{fig:3kall}--\ref{fig:3k5k1k}, we see that
increasing $m_{A}$ has two effects. It suppresses the branching fraction ${\rm BR} (B_s \rightarrow \mu^+ \
\mu^-)$. At the same time it moves the s--channel neutralino annihilation channel to larger values of $M_{1/2}$;
providing larger regions with $0.094 < \abundchi < 0.129$.   However, increasing $m_{A}$ above 1~TeV or so moves
the regions of preferred $\abundchi$ too far to the right, in potential conflict with a lower bound on $m_h$.  In
fact, as can be seen from Fig.~\ref{fig:5kall} (lower right), {\em there is a rough upper bound on $m_{A}$ of
about 1.3 TeV} ({\em and a corresponding lower bound on ${\rm BR} (B_s \rightarrow \mu^+ \ \mu^-) > {\cal O}
(10^{-8})$}) ,~\footnote{The upper bound on $m_A$ will increase slightly as $m_{16}$ increases, but then $m_{16}$
cannot increase significantly without creating a fine-tuning problem.} above which there are no longer any
solutions consistent with the observed $\abundchi$ and the lower bound on the Higgs mass.

It has been shown that, with an integrated luminosity of 15 fb$^{-1}$, experiments at the Tevatron are sensitive
to ${\rm BR} (B_s\rightarrow \mu^+\ \mu^-) > 1.2 \times 10^{-8}$ ~\cite{Arnowitt:2002cq}.  Hence this offers the
prospect of a full exploration of the model in this observable. In addition, the LHC is expected to be sensitive
down to $(3.5 \pm 1.0) \times 10^{-9}$ \cite{Ball:2000ba}. We discuss $BR(B_s \rightarrow \mu^+ \mu^-)$ further
below.

\begin{figure}[t!]
\begin{center}
\begin{minipage}{5in}
\epsfig{file=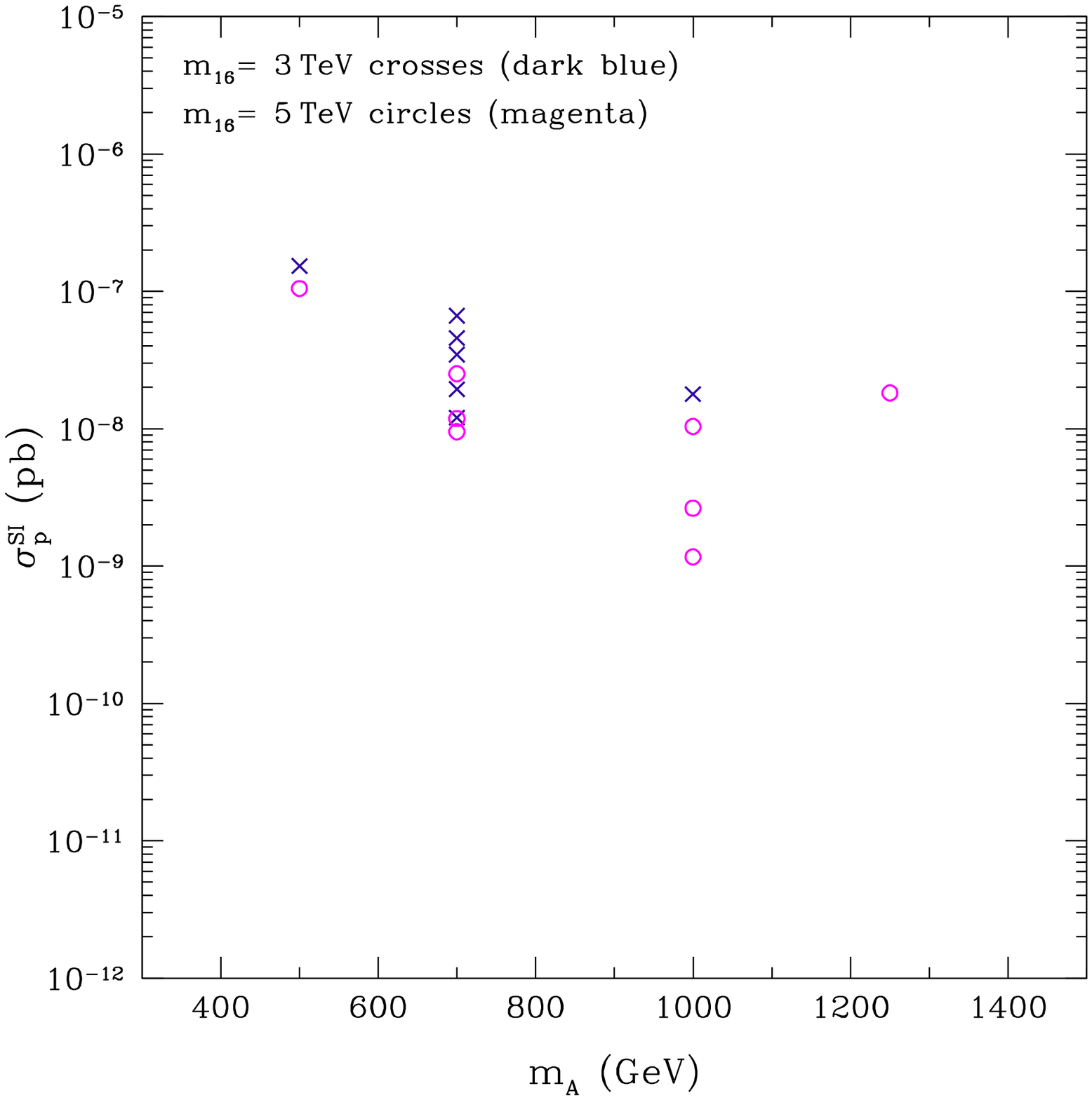,height=5in}
\end{minipage}
\caption{\label{fig:sigp-ma} {\small Spin independent
    neutralino dark matter cross section $\sigsip$ versus $m_A$ for
    $m_{16}= 3$~TeV (dark blue crosses) and 5~TeV (magenta
    circles). The points obey all collider constraints,
    $\chi^2<3$, $0.094< \abundchi < 0.129$ and the new preliminary CDF
    bound ${\rm BR} (B_s \rightarrow \mu^+ \ \mu^-) < 2.0 \times
    10^{-7}$.}}
\end{center}
\end{figure}

\begin{figure}[t!]
\begin{center}
\begin{minipage}{6in}
\epsfig{file=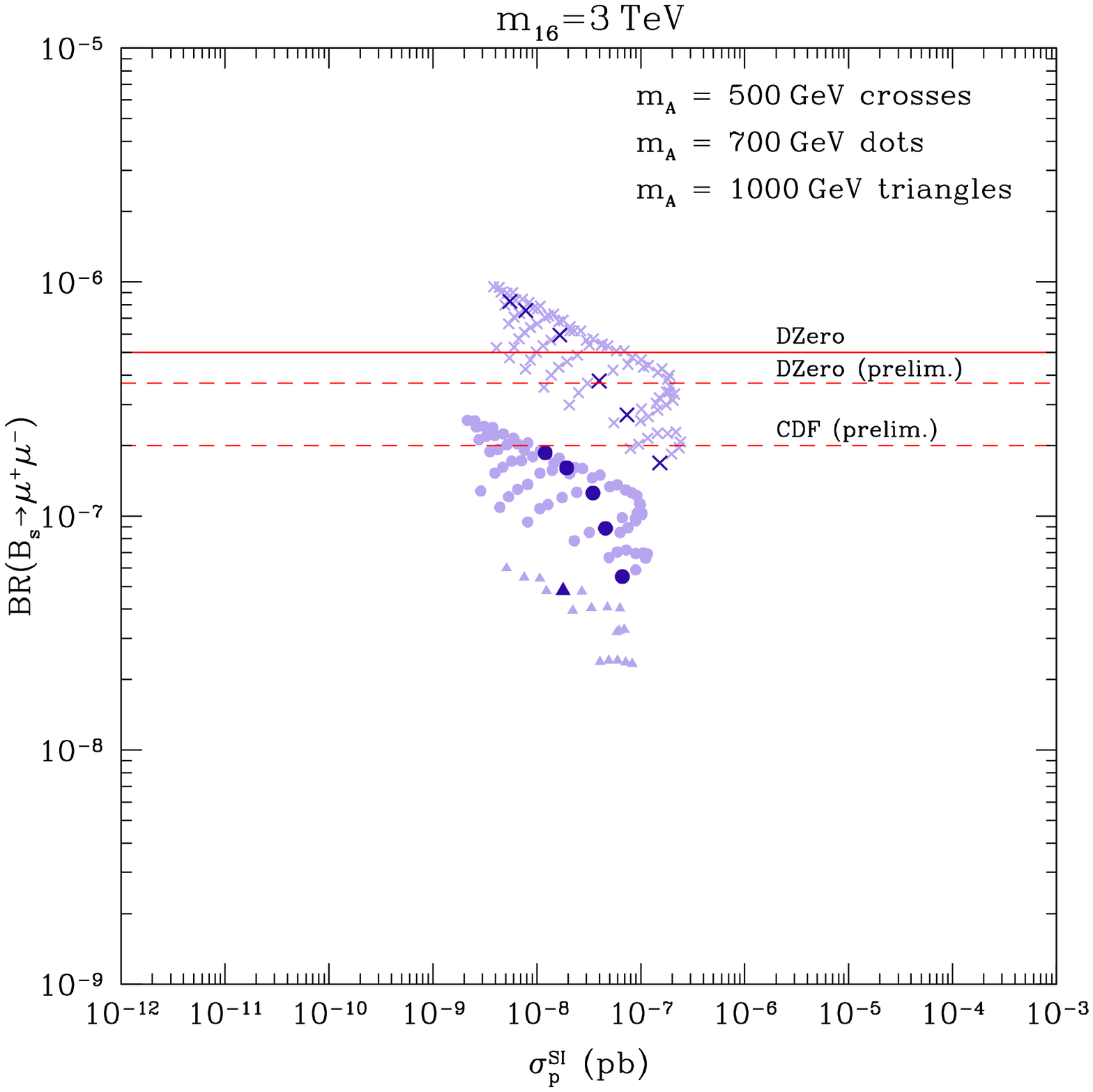,width=3in} \hspace*{-0.15in}
\epsfig{file=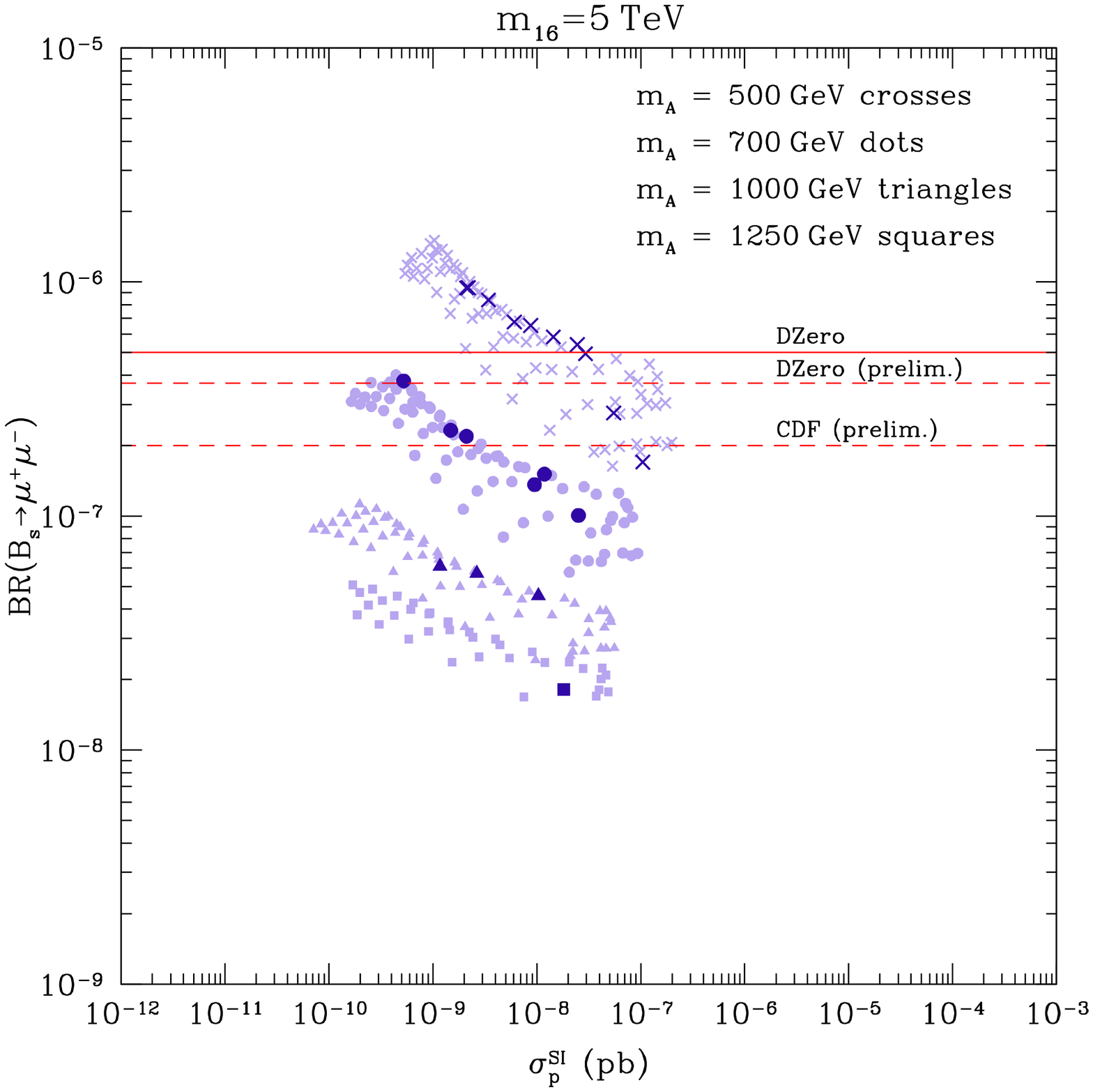,width=3in}
\end{minipage}
\end{center}
\begin{center}
\caption{\label{fig:bsmm-sigp} {\small The branching
ratio ${\rm BR} (B_s \rightarrow \mu^+\ \mu^-)$ versus spin
independent neutralino dark matter cross section $\sigsip$ for
$m_{16} = 3$~TeV (left window) and 5~TeV (right window) and for
several choices of $m_A$. The small light points obey all collider
constraints and $\chi^2 < 3$, while the larger dark points satisfy in
addition $0.094 < \abundchi < 0.129$.} }
\end{center}
\end{figure}

\vspace*{0.1in}
\noindent{\bf \underline{WIMP Direct Detection Search.}}
\hspace*{0.1in} We begin by plotting in Fig.~\ref{fig:sigp-ma}
$\sigsip$, the spin independent neutralino dark matter cross section
relevant for direct dark matter searches, versus $m_A$. We show the
cases $m_{16} = 3$ TeV, $m_A = 500, \ 700, \ 1000$ GeV (dark blue
crosses) and $m_{16} = 5$ TeV, $m_A = 500, \ 700, \ 1000, \ 1250$ GeV
(magenta circles).  The points lie in the preferred region with all
the collider constraints satisfied,
$\chi^2 < 3$ and $0.094 <\abundchi <0.129$.  Like in
Fig.~\ref{fig:bsmm-ma}, for a fixed value of $m_A$, a vertical spread
of points is caused by the changing shape of the cosmologically
favored band in the $\mu, \ M_{1/2}$ plane and also by some variation in
$\tan\beta$ in Figs.~\ref{fig:3kall}--\ref{fig:3k5k1k}. The dominant
contribution to both tree--level and one--loop diagrams typically
comes from a $t$--channel exchange of the heavier scalar Higgs boson
$H^0$. Thus one would normally expect $\sigsip\sim m_{A}^{-4}$ since
$m_A\simeq m_H$. However, the coupling $\chi \chi H^0$ is sensitive to
the bino/higgsino composition of the lightest neutralino and vanishes
in the limit of the neutralino becoming one of these two states.  The
bino fraction increases along the cosmologically favored
band from as low as some $70\%$ at low
$\mu$ to almost $100\%$ at largest allowed $\mu$. This causes
$\sigsip$ to decrease with $\mu$ even for a fixed $m_A$.

Since both $BR(B_s \rightarrow \mu^+ \mu^-)$ and $\sigsip$ are to a large
degree determined by $m_A$, but also by the LSP composition,
one finds interesting correlations which are shown in
Fig.~\ref{fig:bsmm-sigp}. The small light points obey all collider constraints
and $\chi^2 < 3$, while the larger dark points satisfy in addition
$0.094 < \abundchi < 0.129$.
For fixed $m_A$ and $M_{1/2}$, $\sigsip$ decreases with increasing
$\mu$ due to the increasing bino content of the neutralino.
On the other hand ${\rm BR} (B_s \rightarrow \mu^+ \ \mu^-)$
increases as $\mu$ increases, since the flavor violating vertex correction
is proportional to $\mu$ (see below).
Hence, as seen from Fig.~\ref{fig:bsmm-sigp}, in the allowed parameter space
of the MSO$_{10}$SM, the values for
$\sigsip$ and ${\rm BR} (B_s \rightarrow \mu^+ \ \mu^-)$
are (as one moves along the cosmologically favored band,
for a fixed $m_A$) {\em inversely} correlated.
This behavior sharply contrasts with the direct correlation
between the two quantities in the CMSSM and its extensions presented
in~\cite{bkk05}, where, with the usual parametrization,
the masses of the Higgs bosons grow along the cosmologically favored band.
Thus, an improved limit on $BR(B_s \rightarrow \mu^+ \mu^-)$ does not need
to necessarily imply less promising prospects for measuring
$\sigsip$, and vice versa.
In addition, in Fig.~\ref{fig:bsmm-sigp} we see that,
comparing the case of $m_{16} = 3$ and 5 TeV (again, for fixed $m_A$)
in the latter case the range of $\sigsip$ for the (dark) points in the
cosmologically preferred region extends to larger values of
${\rm BR} (B_s \rightarrow \mu^+ \ \mu^-)$. This is
because the flavor violating vertex of the CP odd Higgs scales
roughly like $\mu A_t/m_{16}^2$ and for larger
values of $m_{16}$ the acceptable regions of parameter space extend to even
larger values of $\mu$.

\begin{figure}[t!]
\begin{center}
\hspace*{.5in}
\begin{minipage}{5in}
\epsfig{file=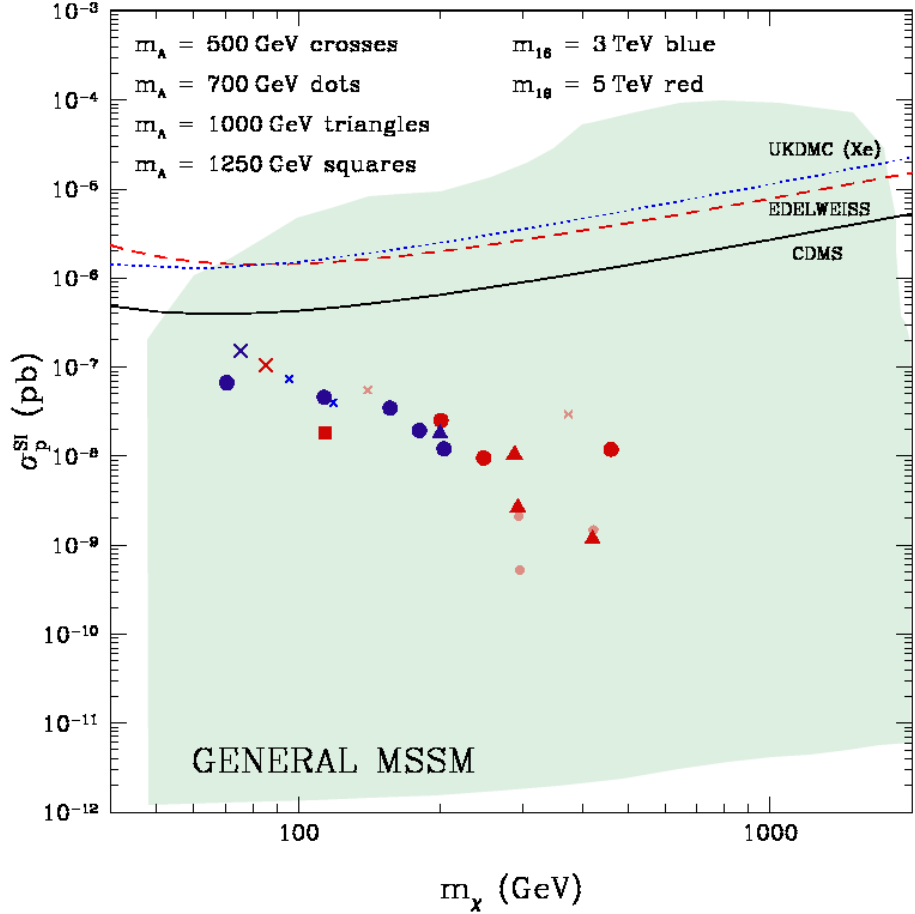,height=5in}
\end{minipage}
\caption{\label{fig:sigp-mx} {\small Spin independent dark matter
    cross section $\sigsip$ \vs\ the neutralino mass $\mchi$ for
    several choices of $m_{16}$ and $m_A$. Big points obey all
    collider constraints, $\chi^2<3$, $0.094< \abundchi < 0.129$ and
    the new preliminary CDF bound ${\rm BR} (B_s \rightarrow \mu^+ \
    \mu^-) < 2.0 \times 10^{-7}$ while for the small, light points the
    last constraint is relaxed to ${\rm BR} (B_s \rightarrow \mu^+ \
    \mu^-) < 5.0 \times 10^{-7}$.  The lightly shaded background
    represents predictions in the general MSSM for which the bound
    from ${\rm BR} (B_s \rightarrow \mu^+ \ \mu^-)$ was not applied.}}
\end{center}
\end{figure}

In Fig.~\ref{fig:sigp-mx} we plot $\sigsip$ versus the neutralino mass.
We show the same cases of $m_{16}$ and $m_A$
as in Figs.~\ref{fig:bsmm-ma}--\ref{fig:bsmm-sigp}.
Big points obey all collider constraints, $\chi^2<3$,
$0.094< \abundchi < 0.129$ and the new preliminary CDF bound
${\rm BR} (B_s \rightarrow \mu^+ \ \mu^-) < 2.0 \times 10^{-7}$
while for the small, light points the last constraint is relaxed
to ${\rm BR} (B_s \rightarrow \mu^+ \ \mu^-) < 5.0 \times 10^{-7}$.
In the (light blue) background we plot predictions obtained in the
general MSSM by scanning large ranges of parameters, including $\tanb$
from 10 up to 65~\cite{knrr1} but do not apply the bound from
${\rm BR} (B_s \rightarrow \mu^+ \ \mu^-)$. Note, most regions of parameter
space should be observable as dark matter search experiments are expected
to probe down to values of $\sigsip \gtrsim 10^{-8}$ pb
with current/upgraded detectors within the next year or so (\eg, by CDMS).
Future one--tonne detectors are planned to reach $\sigsip \gtrsim 10^{-10}$
pb, thus probing the entire favored parameter space of the MSO$_{10}$SM.


\vspace*{0.1in} \noindent{\bf\underline{WIMP Indirect Detection
Search.}}\hspace*{0.1in} Apart from direct detection, dark matter WIMP
signals can be detected in a number of indirect detection search
channels. Here we compute detection rates for a high--energy neutrino
flux from the Sun, for gamma rays from the Galactic center, and for
antiprotons and positrons from the Galactic halo, by applying
DARKSUSY~\cite{darksusy} to some popular Galactic halo models.  As we
will see, prospects for WIMP detection in these modes are generally
less promising than in direct detection experiments, although a
sizeable number of cosmologically preferred cases can still be
probed. On top of this, astrophysical uncertainties are often
sizeable, especially for the latter two modes since propagation of
antiprotons and positrons in the Milky Way is poorly understood.

\begin{figure}[t!]
\begin{center}
\begin{minipage}{6in}
\epsfig{file=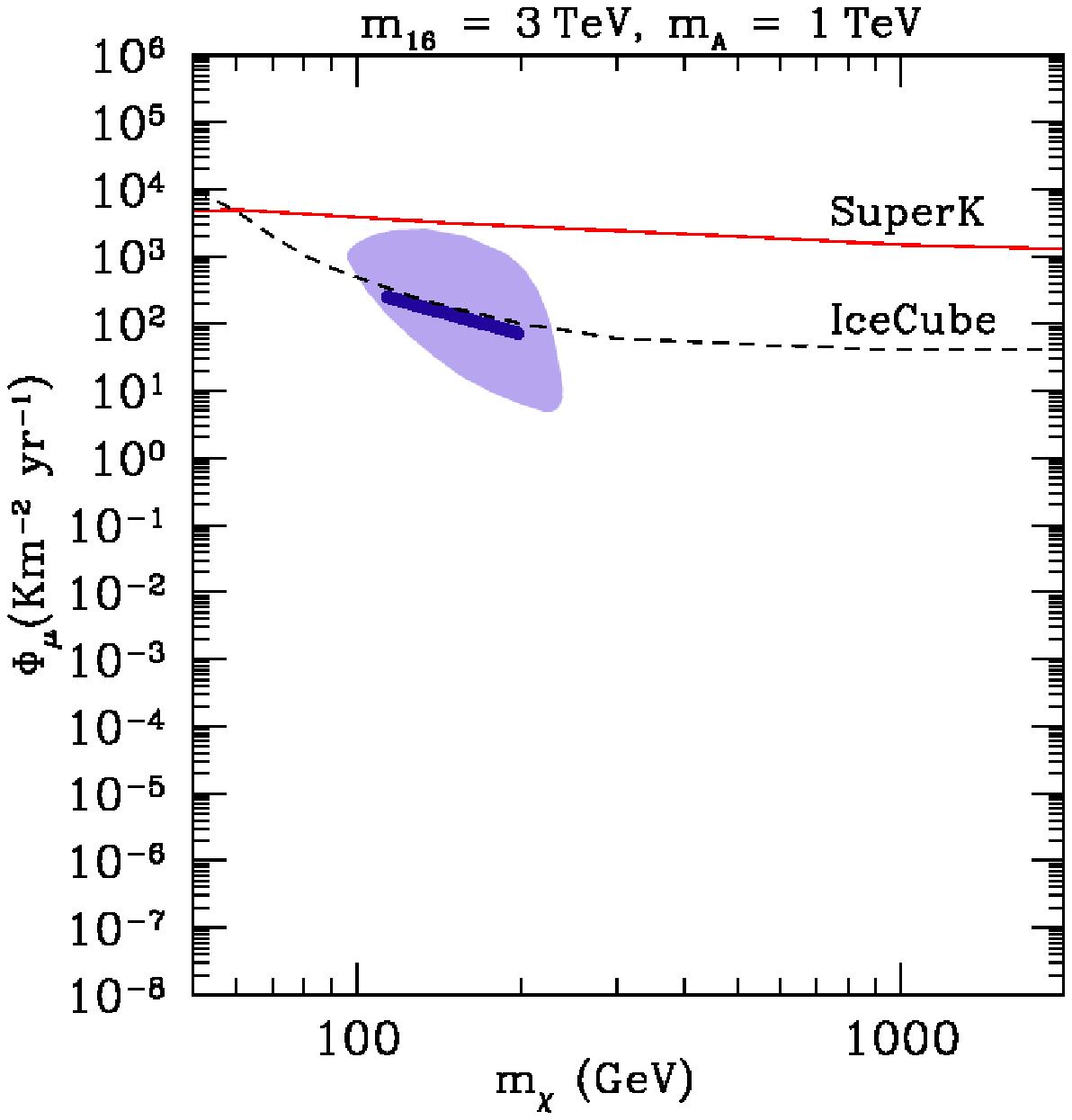,width=3in} \hspace*{-0.15in}
\epsfig{file=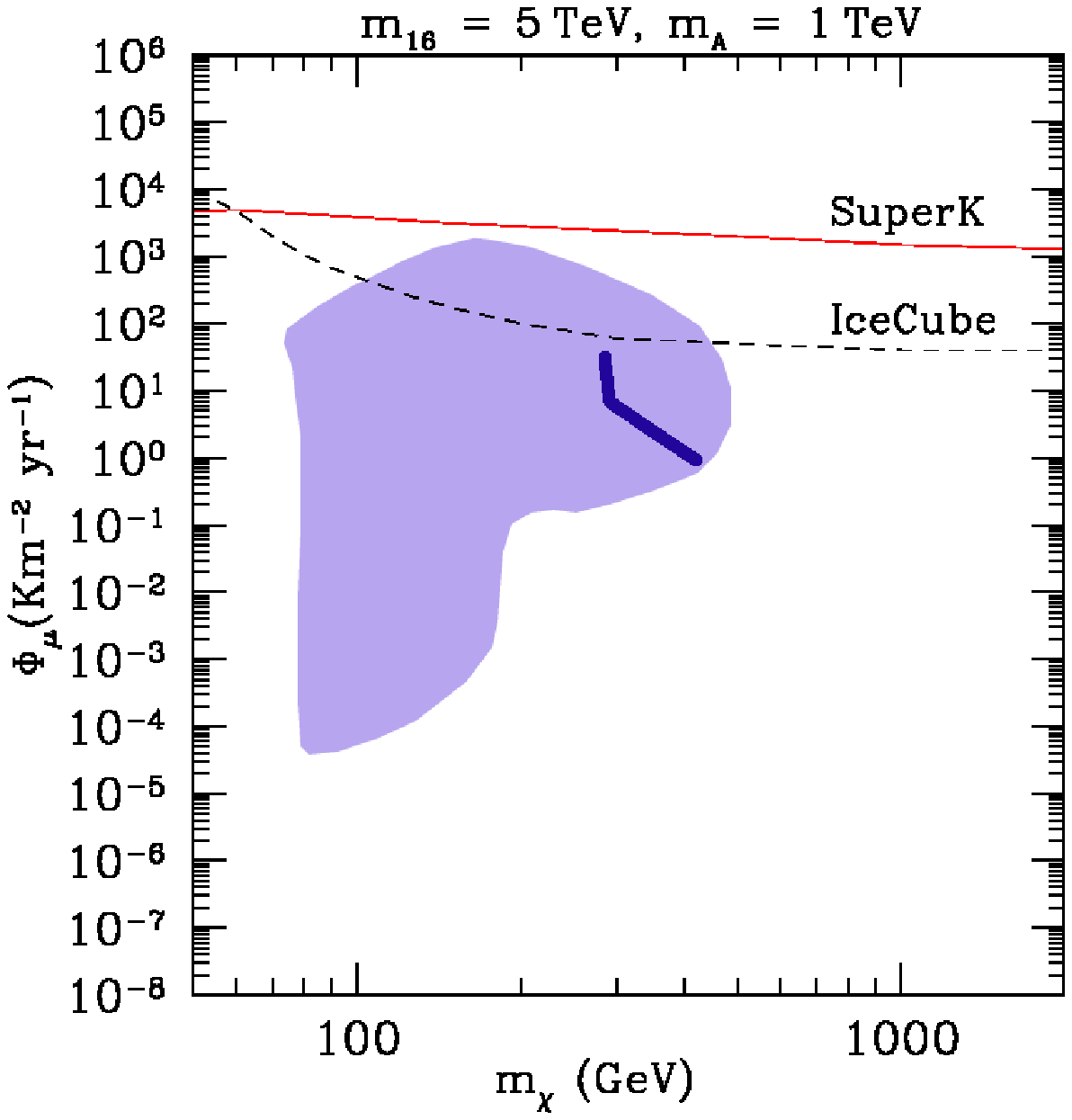,width=3in}
\end{minipage}
\end{center}
\begin{center}
\caption{\label{fig:sunmuflux} {\small Muon flux from the Sun  \vs\
    the neutralino mass $\mchi$ for
    $m_A=1$~TeV and $m_{16}=3$~TeV (left window) and $m_{16}=5$~TeV
    (right window) assuming the energy threshold $E_{\mu}^{\rm th}=
    25$~GeV. In light blue (shaded) areas the collider constraints and
    $\chi^2<3$ have been applied but not the cosmological constraint
    $0.094< \abundchi < 0.129$ nor the bound on ${\rm BR} (B_s
    \rightarrow \mu^+ \ \mu^-)$. Points along the dark blue line
    additionally satisfy the cosmologically preferred range $0.094 <
    \abundchi < 0.129$. The current limit from SuperKamiokande is
    marked as a red solid line while future reach of IceCube is marked
    as a black dashed line.} }
\end{center}
\end{figure}

Instead of undertaking a full presentation of all the cases discussed above,
we will select a representative value of $m_A=1$~TeV and will present our
results assuming all collider constraints (except for the ${\rm BR}
(B_s \rightarrow \mu^+ \ \mu^-)$ bound which, for the chosen value of $m_A$,
is irrelevant), $\chi^2<3$ and $\abundchi < 0.129$, and then show the
(typically strong) impact of imposing the cosmological constraint
$0.094< \abundchi < 0.129$.
Whenever the WIMP relic density is too small ($\abundchi < 0.094$),
we will not rescale the rates.

Halo WIMPs pass through the Sun and other celestial bodies and occasionally
become gravitationally trapped inside their cores.
Once enough of them have accumulated, they start pair--annihilating into pairs
of SM particles ($q\bar q$, $WW$, $ZZ$, \etc), which subsequently decay
via two-- and three--body processes, like
$b\rightarrow cl\nu_l$, $W\rightarrow {\bar\nu} \nu$, \etc.
Among the decay products only neutrinos can escape out of the Sun's
core. A muon neutrino from the Sun, when passing through the Earth,
may produce a muon which will generate a {\v C}erenkov shower in a nearby
under--ice or under--water detector.

The calculation of the WIMP capture rate and subsequent annihilation,
as well as propagation of decay products is
rather well understood~\cite{jkg96} and fairly insensitive to the choice
of a Galactic halo model.
(We assume the local dark matter density $\rho_0=0.3~{\rm GeV}/{\rm cm}^3$
and a Maxwellian velocity distribution with the peak velocity $v_0=270$~km/s.)
The capture rate in the Sun, typically determined by spin dependent
interactions involving $Z$--boson exchange, is basically independent of
$m_A$. On the other hand, the shape of the cosmologically favored band
in the $\mu, \ M_{1/2}$ plane does depend on $m_A$.

In Fig.~\ref{fig:sunmuflux} we plot the high--energy muon flux from the Sun
for $m_A=1$~TeV, and $m_{16}=3$~TeV (left window) and
$m_{16}=5$~TeV (right window).
The IceCube energy threshold $E_{\mu}^{\rm th}= 25$~GeV has been applied.
In light blue (shaded) areas the collider constraints and $\chi^2<3$ have been
applied but not the cosmological constraint $0.094< \abundchi < 0.129$
nor explicitly the bound on ${\rm BR} (B_s \rightarrow \mu^+ \ \mu^-)$.
Dark lines correspond to points which additionally satisfy the
cosmologically preferred range $0.094 < \abundchi < 0.129$.
We can see that the favored configurations fall below the current best limit
from Super-Kamiokande~\cite{superk} and are on a bordeline for being probed at
IceCube~\cite{icecube}.

\begin{figure}[t!]
\begin{center}
\begin{minipage}{6in}
\epsfig{file=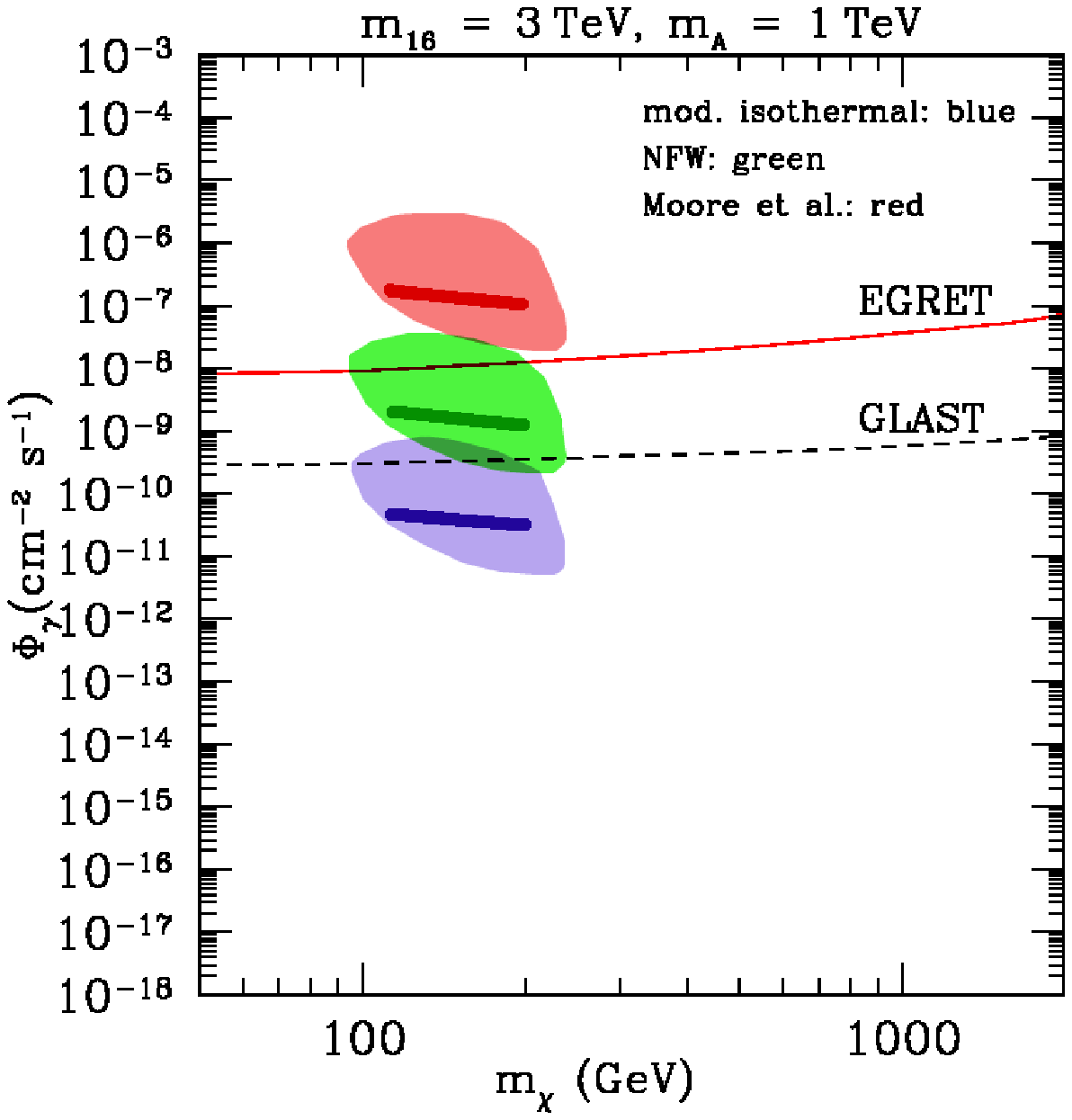,width=3in} \hspace*{-0.15in}
\epsfig{file=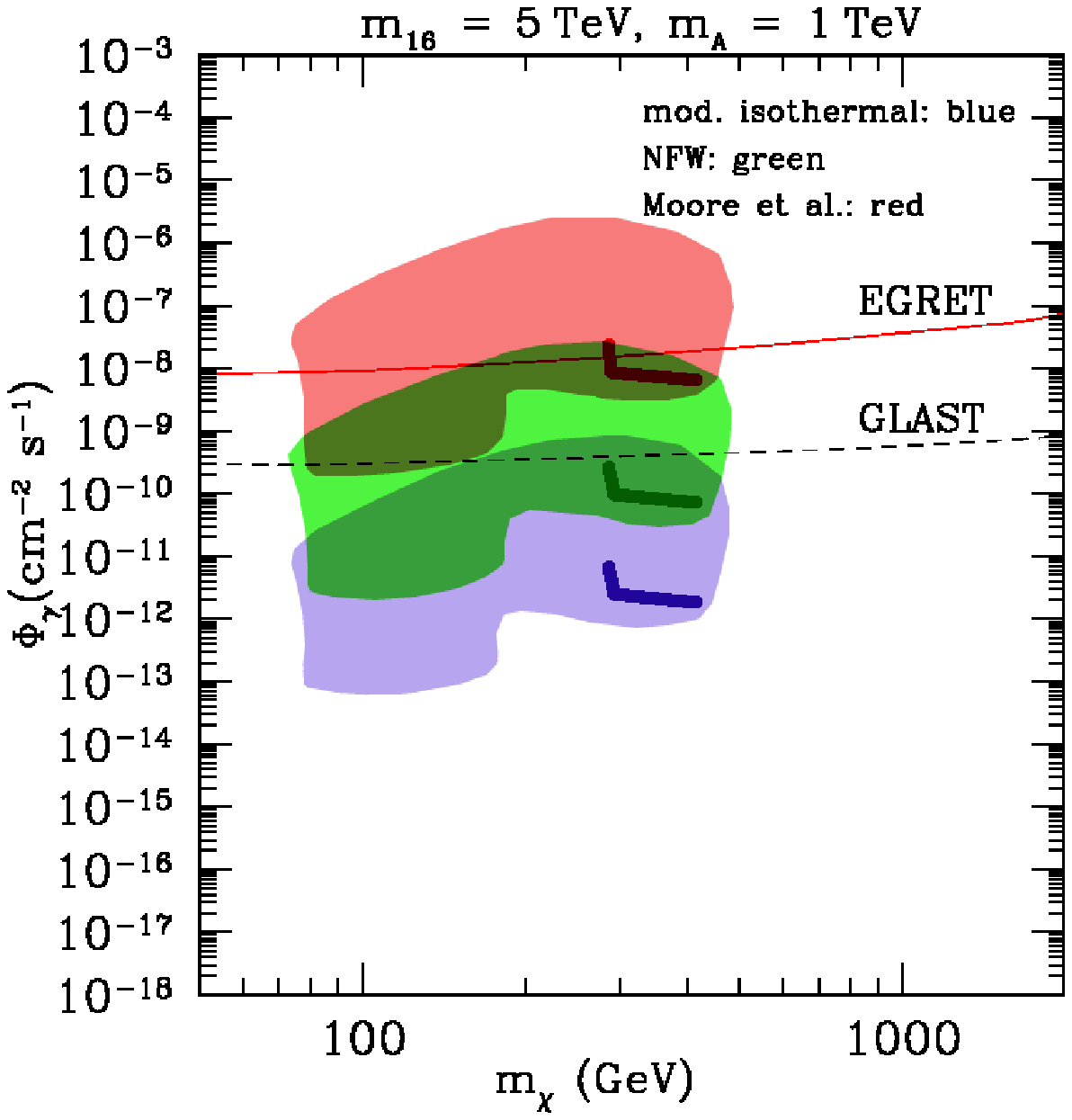,width=3in}
\end{minipage}
\end{center}
\begin{center}
\caption{\label{fig:phigamflux} {\small Gamma ray fluxes from the
    Galactic center  \vs\ the neutralino mass $\mchi$ for $m_A=1$~TeV
    and for $m_{16}=3$~TeV (left window)
    and $m_{16}=5$~TeV (right window), assuming the energy threshold
    $E_{\gamma}^{\rm th}= 1$~GeV, as in GLAST. Results for three different halo
    models are shown, as
    described in the text. In light-colored (shaded) areas
    collider constraints and $\chi^2<3$ have been applied but not the
    cosmological constraint $0.094< \abundchi < 0.129$ nor the bound on
    ${\rm BR} (B_s \rightarrow \mu^+ \ \mu^-)$. Dark lines
    corresponding to each halo model mark points which additionally satisfy the
    cosmological constraint $0.094 < \abundchi < 0.129$. The current
    limit from EGRET is marked as a red solid curve while an expected
    reach of GLAST is marked as a black dashed line.}}
\end{center}
\end{figure}

Predictions for the fluxes of gamma rays from the Galactic center and
antiprotons and (to a lesser extent) positrons from the Galactic
halo depend on the assumed halo model. Several profiles of dark
matter distribution have been discussed in the literature, many of which
can be parametrized by

\begin{equation}
\rho(r)= \rho_0 \frac{(r/r_0)^{-\gamma}}
            {\left[1+\left(r/a\right)^\alpha\right]^{\frac{\beta-\gamma}{\alpha}}}
             \left[1+(r_0/a)^\alpha\right]^{\frac{\beta-\gamma}{\alpha}},
\end{equation}
where $\rho(r)$ is the radial dependence of the halo WIMP density,
$\rho_0=0.3~{\rm GeV}/{\rm cm}^3$ is the local
dark matter density, $r_0$ is our distance to the Galactic center
and $a$ is a distance scale.

Here we consider three distinct and popular choices:
\begin{itemize}

\item a sperically symmetric modified isothermal profile~\cite{isothermalmodel}, for
  which $(\alpha, \beta, \gamma)= (2, 2, 0)$, $r_0=8.5$~kpc  and $a=3.5$~kpc;

\item the Navarro, Frenk and White (NFW) profile~\cite{nfwhalo95}, for
  which $(\alpha, \beta, \gamma)= (1, 3, 1)$, $r_0=8.0$~kpc  and $a=20$~kpc;

\item the Moore, \etal, profile~\cite{morehalo99}, for
  which $(\alpha, \beta, \gamma)= (1.5, 3, 1.5)$, $r_0=8.0$~kpc and $a=28$~kpc.

\end{itemize}

Towards the Galactic center the halo density is expected to be larger
than in our local neighborhood but otherwise constant in the case of
the modified isothermal model, or divergent as $\rho(r)\sim 1/r$ (NFW)
and $\rho(r)\sim 1/r^{1.5}$ (Moore, \etal). It is therefore natural to
expect enhanced neutralino WIMP annihilations in the core of the
Galactic center. Among annihilation products, high--energy gamma rays
are unique in that they point back directly to the Galactic
center. The flux is proportional to the DM number density squared
integrated along the line of sight.  Monochromatic photons with energy
$E_\gamma\simeq\mchi$ would provide a spectacular signal but the rates
are low compared to the diffuse gamma radiation from cascade decays of
WIMP annihilation products.

\begin{figure}[b!]
\begin{center}
\begin{minipage}{6in}
\epsfig{file=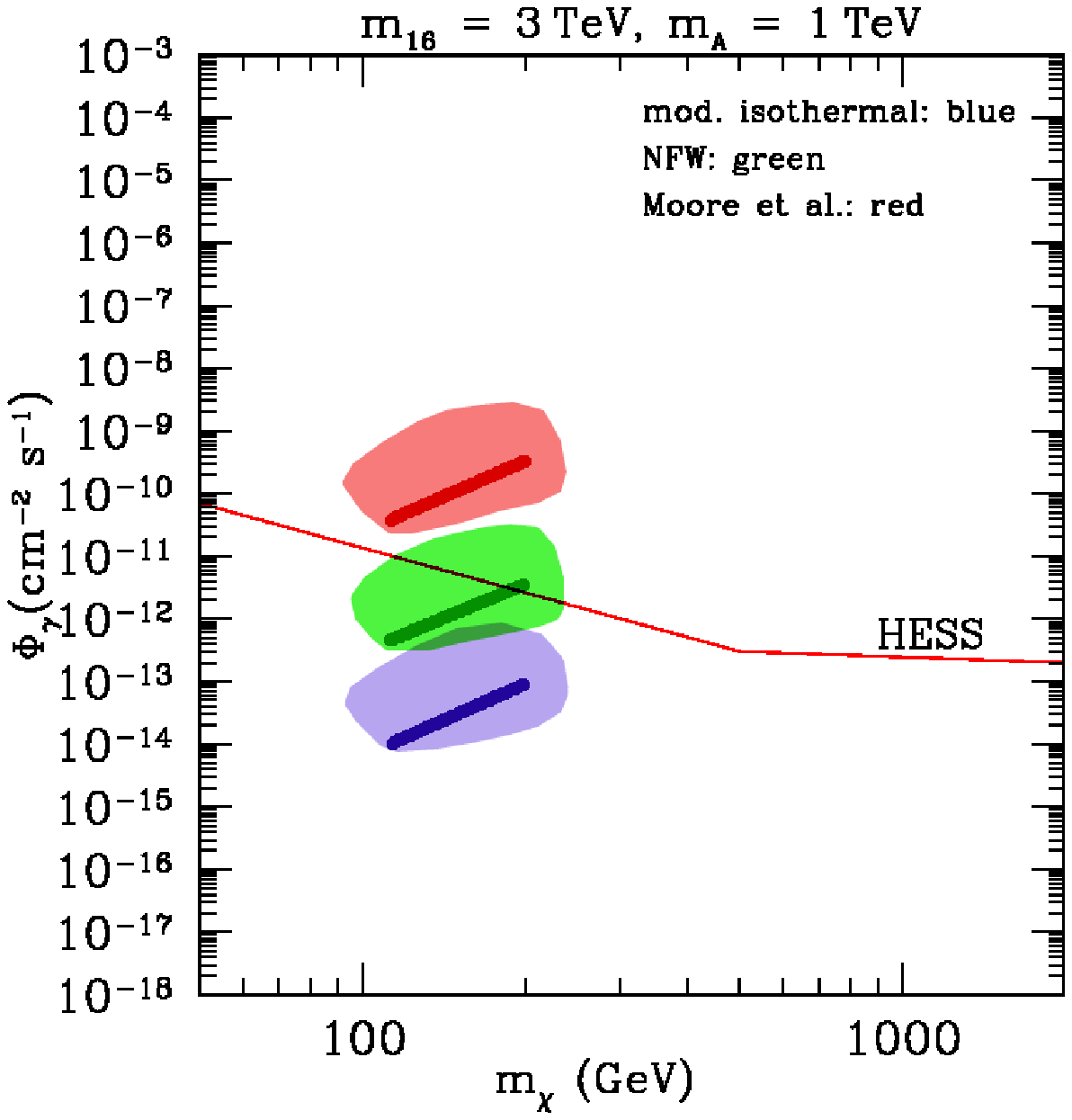,width=3in} \hspace*{-0.15in}
\epsfig{file=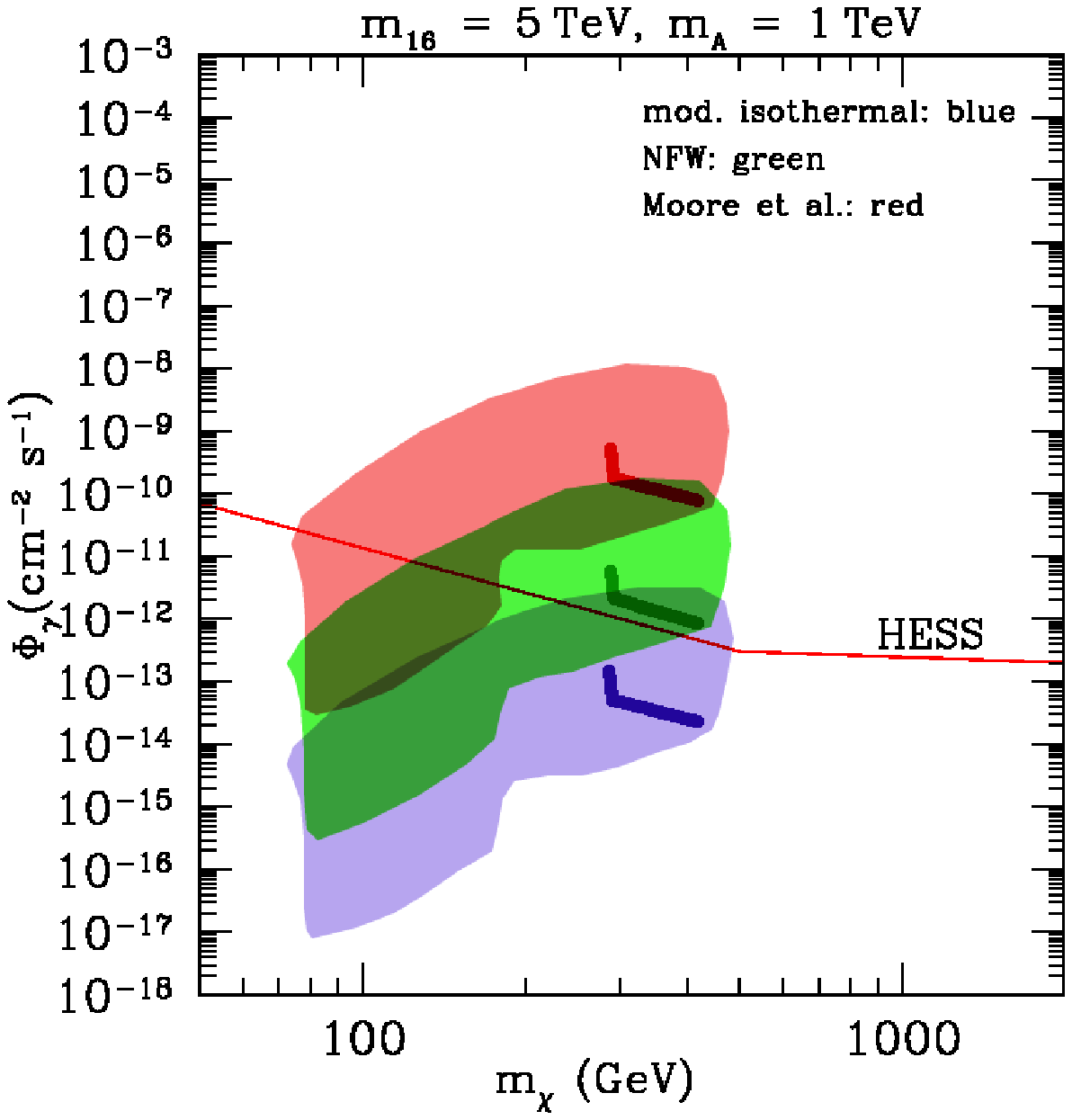,width=3in}
\end{minipage}
\end{center}
\begin{center}
\caption{\label{fig:phigamflux-hess} {\small Same as
    Fig.~\protect\ref{fig:sunmuflux} but assuming the energy threshold
    $E_{\gamma}^{\rm th}= 60$~GeV, as in HESS whose limit is marked as
    a red solid line.}}
\end{center}
\end{figure}

In Fig.~\ref{fig:phigamflux} we plot the diffuse high--energy gamma flux from
the Galactic center, integrated over the cone of $0.001$~sr.
We apply the continuous gamma energy threshold $E_{\gamma}^{\rm th}= 1$~GeV,
as planned for GLAST~\cite{bbmn04}.
We also denote an upper limit from EGRET~\cite{hooper+dingus05}
and the expected reach of GLAST after three years of taking data.
The rates are independent of the choice of $m_A$. Depending on the
choice of the halo model, GLAST may have a good chance of detecting a
WIMP signal.

For comparison, in Fig.~\ref{fig:phigamflux-hess} we plot the same quantity
but apply the larger energy threshold of 60~GeV, typical of HESS.
For the case $m_{16}=5$~TeV the cosmologically favored values of $\mchi$
are higher than for $m_{16}=3$~TeV, and accordingly the resulting diffuse
gamma radiation will be more accessible to HESS~\cite{hess} than to GLAST.
Thus HESS has a good chance of detecting a signal in the region
of the parameter space which is to
some extent complementary to the reach of GLAST.

\begin{figure}[t!]
\begin{center}
\begin{minipage}{6in}
\epsfig{file=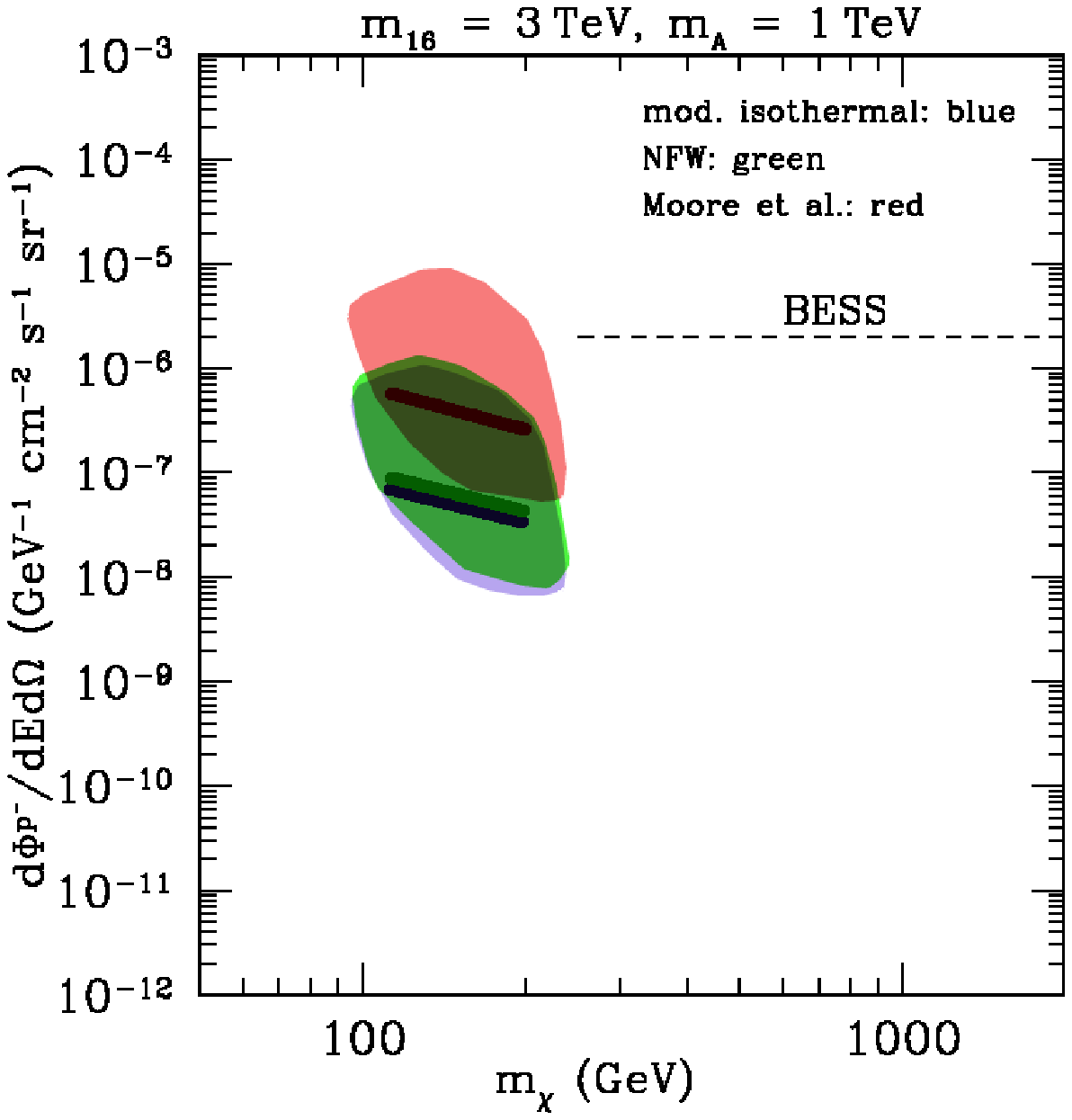,width=3in} \hspace*{-0.15in}
\epsfig{file=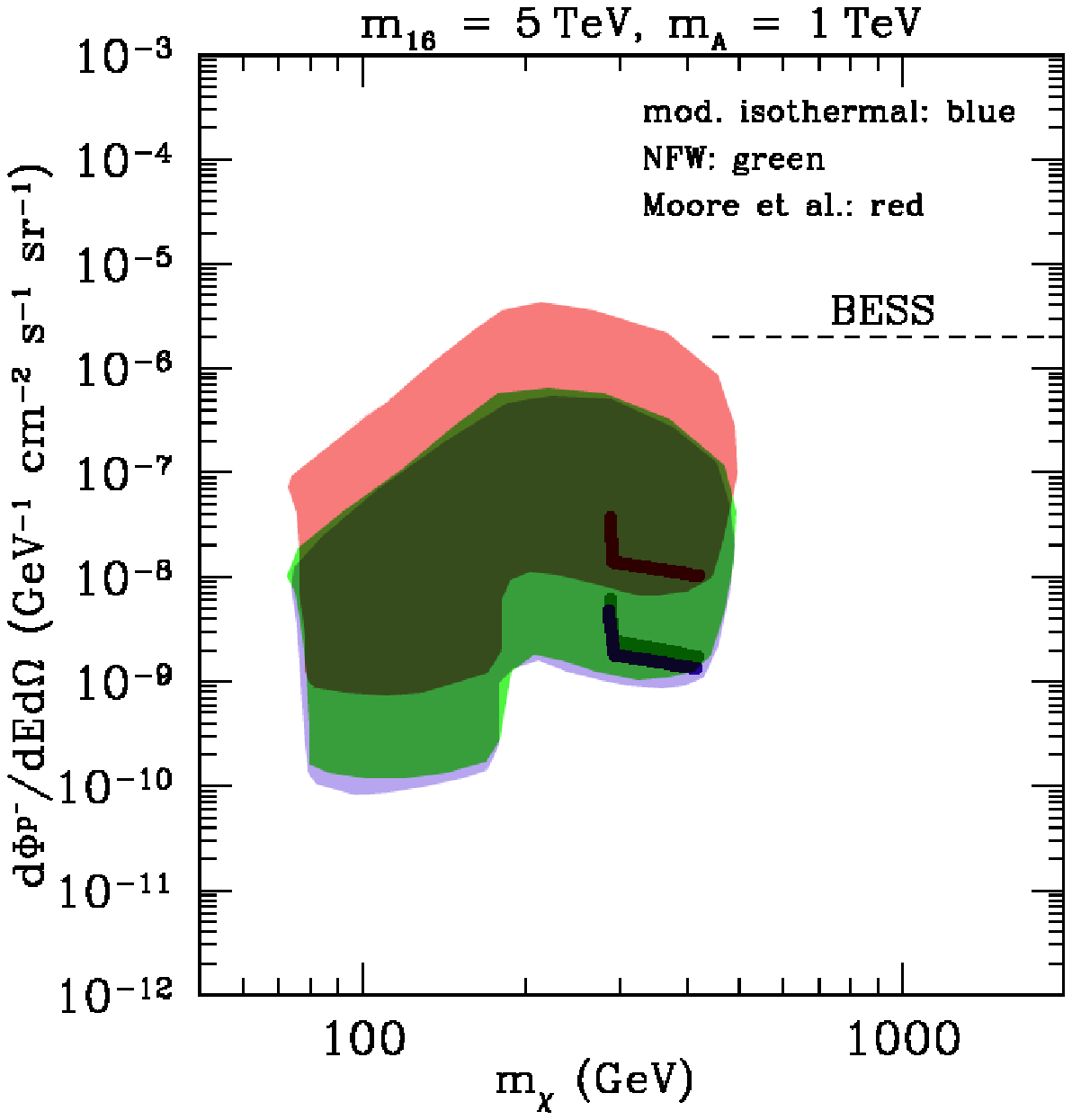,width=3in}
\end{minipage}
\end{center}
\begin{center}
\caption{\label{fig:phiapflux} {\small Same as
    Fig.~{\protect\ref{fig:sunmuflux}} but for antiproton fluxes from
    the Galactic halo. A sensitivity of BESS is indicated with a
    dashed line.}}
\end{center}
\end{figure}

Another way of looking for WIMPs is to look for an excess of antiparticles
in cosmic rays originating from WIMP annihilation in the Galactic halo,
despite large uncertainties~\cite{jkg96,beu99}. A flux of antiprotons from
astrophysical sources is expected to fall off dramatically at low energies,
unlike that from WIMP annihilation in the halo.
For a detailed discussion of antiproton fluxes from WIMP annihilation
and from background sources see Ref.~\cite{beu99}.

In Fig.~\ref{fig:phiapflux} we plot a differential antiproton flux
$d\,\Phi^{p^-}/d E\, d\Omega$ \vs\ $\mchi$.
We evaluate the quantity assuming the peak of the flux at
$E_{p^-} \sim 1.76$~GeV to coincide with a peak in the
kinetic energy distribution of the BESS experiment~\cite{bess}.
The plots do not depend on $m_A$ and the halo model dependence is fairly
weak, except for the Moore, \etal, model for which the rate is somewhat
higher. As we can see, the cosmologically favored points fall below
the BESS sensitivity.

\begin{figure}[t!]
\begin{center}
\begin{minipage}{6in}
\epsfig{file=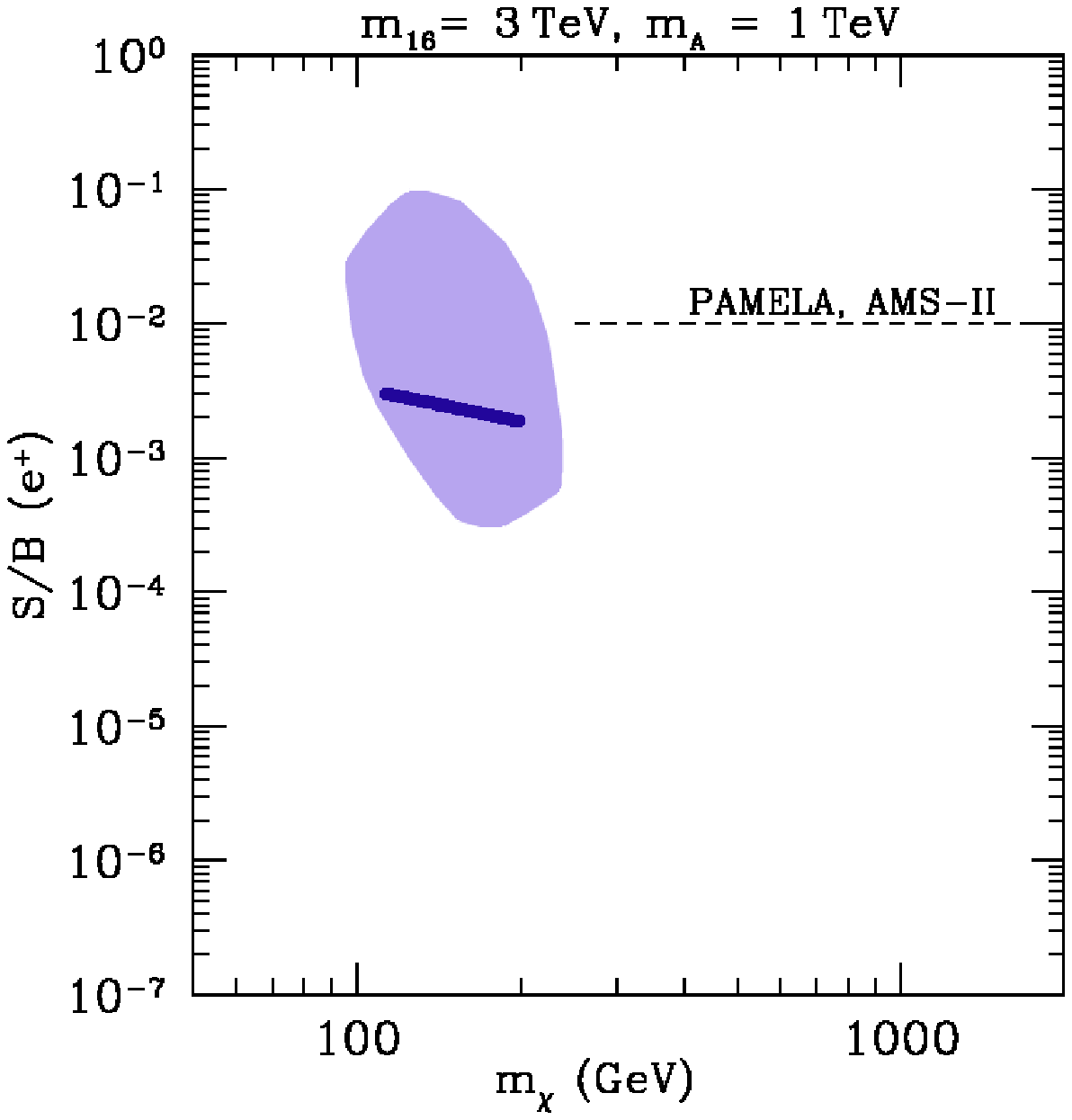,width=3in} \hspace*{-0.15in}
\epsfig{file=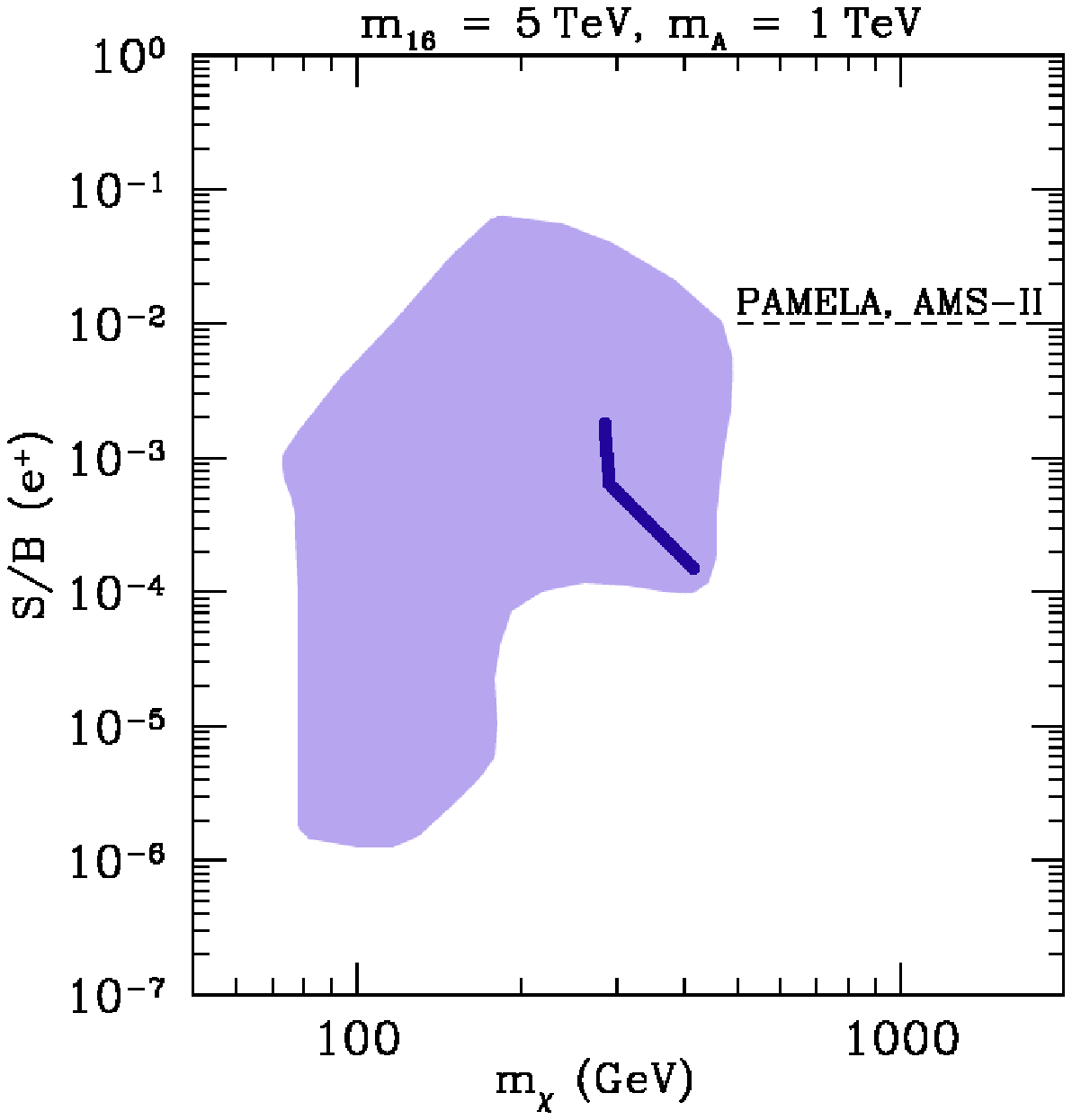,width=3in}
\end{minipage}
\end{center}
\begin{center}
\caption{\label{fig:phiaeflux} {\small Same as
    Fig.~{\protect\ref{fig:sunmuflux}} but for positron flux from WIMP
    annihilation in the Galactic halo presented as
    signal--to--background ratio B/S, as described in the
    text. Rates as low as $0.01$ may be detectable at PAMELA and
    AMS--II.}}
\end{center}
\end{figure}

Finally, high energy positrons may be a signature of WIMP annihilation in the
halo. Again, like in the case of antiprotons, background from astrophysical
processes (mostly spallation of cosmic rays off the interstellar medium)
and other astrophysical uncertainties are rather large.
Production of hard positrons in $e^+e^-$ pairs
from direct WIMP annihilation is tiny due to helicity suppression.
On the other hand, annihilation into $WW$ and $ZZ$ pairs,
followed by their decays into $e^+e^-$ pairs will result
in a positron flux peaked at an energy of $\sim\mchi/2$.
This feature gives some hope of distinguishing the signal (S) due to
WIMP annihilation from the background (B) which is typically two to
three orders of magnitude larger.

In Fig.~\ref{fig:phiaeflux} we plot the signal--to--background ratio S/B.
The positron flux is evaluated at an optimal energy of $\mchi/2$ and
is compared with background flux, as parametrized by Feng, \etal~\cite{fmw01}.
Halo model dependence is negligible since energetic positrons produced
in the solar neighborhood would be detectable.
A sensitivity of positron search experiments PAMELA~\cite{pamela}
and AMS--II~\cite{ams} is indicated where S/B rates as low as $0.01$
may be detectable.
As can be seen from the figure, this is unlikely to be sufficient to probe our
predicted rates for cosmologically favored regions.


\subsection{Some representative points}

In Table 1 we present the input parameters and resulting Higgs and SUSY
spectra for four representative points in
SUSY parameter space which are consistent with all available data.
We also present several other relevant experimental observables,
including the SUSY contribution to the anomalous magnetic moment of the muon,
$a_\mu^{SUSY}$, and the branching ratio for the process $B \rightarrow X_s
\gamma$, for these points.
Note that one stop and/or one stau, and the charginos and neutralinos
are relatively light.
Also, as might be expected with heavy first and second generation sleptons,
the value of $a_\mu^{SUSY}$ is small (of order 1 - 3 $\times 10^{-10}$).
Finally, although ${\rm BR}(B \rightarrow X_s \gamma)$ was not included in
$\chi^2$, and the predicted values are typically outside the experimentally
allowed range, it is not difficult to find
consistent solutions to all the data, including this observable.

\begin{table}
\label{t:table} \caption{The input parameters, $\chi^2$ fits and Higgs and
SUSY spectra for four representative
points consistent with all available data.}
$$
\begin{array}
{|l|c|c|c|c|c|} \hline
{\rm Data \; points}  & & 1 & 2 & 3 & 4 \\
\hline
 {\rm Input\; parameters}  & &  &  &  & \\
\hline
\;\;\;\alpha_{G}^{-1} & & 24.86 & 24.90  & 25.01 & 25.30 \\
\;\;\; M_G \times 10^{-16} & & 2.96  & 2.82  & 2.59 & 2.30  \\
\;\;\;\epsilon_3 & & -0.036   & -0.036    & -0.033   & -0.029  \\
\;\;\;\lambda  & &  0.71   &  0.72  & 0.69   & 0.73  \\
\hline
\;\;\; m_{16} & & 3000 & 3000  & 3000  & 5000\\
\;\;\; m_{10}/m_{16} & & 1.33  & 1.35   & 1.37   & 1.35  \\
\;\;\; \Delta m_H2  & & 0.15   & 0.17   & 0.16  & 0.16  \\
\;\;\; M_{1/2} & & 330 &  380 &  500 & 700 \\
\;\;\;\mu & & 260 &  240 &  360  & 400 \\
\;\;\; \tan\beta & & 51.93  &  51.97 &  50.7 & 51.61  \\
\;\;\; A_0/m_{16} &  & -1.83 &  -1.83 &  -1.91  & -1.88   \\
\hline \hline
\chi^2 \; {\rm observables} & {\rm Exp}\;(\sigma)  &    &  &  &  \\
\hline
\;\;\;M_Z & 91.188 \;(0.091) & 91.19   & 91.19  & 91.19   & 91.22 \\
\;\;\;M_W & 80.419 \;(0.080) & 80.42   & 80.41   &  80.42  & 80.40  \\
\;\;\;G_{\mu}\times 105 & 1.1664 \;(0.0012) & 1.166  & 1.166   & 1.166  & 1.166 \\
\;\;\;\alpha_{EM}^{-1} & 137.04\; (0.14) & 137.0  & 137.0   & 137.0   & 137.0  \\
\;\;\;\alpha_s(M_Z)  & 0.1172 \; (0.0020) & 0.1171   & 0.1169 & 0.1171  & 0.1173 \\
\;\;\;\rho_{new}\times 103 & -0.200 \; (1.10) & 0.363   & 0.287    & 0.442  & -0.332 \\
\hline
\;\;\;M_t   & 178 \; (4.3)    & 176.3 &  176.7 &  175.9  & 177.5 \\
\;\;\;m_b(m_b) & 4.20 \; (0.20)  &  4.27  &  4.3 &  4.27   & 4.19 \\
\;\;\;M_{\tau} & 1.7770 \; ( 0.0018)  & 1.777 &  1.777 &  1.777  & 1.777 \\
\hline
 {\rm TOTAL}\;\;\;\; \chi^2 & & 0.56 & 0.57 & 0.73  & 0.18 \\
\hline \hline
\;\;\; h      &        &  121  & 121  & 118  &  119.5 \\
\;\;\; H       &        &  559   & 791 & 789 & 1139 \\
\;\;\; A        &       &  500  & 700  & 700 & 1000 \\
\;\;\; H^+         &      &  541  & 758 & 758 & 1083 \\
\;\;\; \chi0_1 &  &  131  & 148  & 204 & 287 \\
\;\;\; \chi0_2  & &  217  & 221  & 331 & 393 \\
\;\;\; \chi^+_1  & &   214  & 214 & 328 & 387 \\
\;\;\; \tilde g         &  &  854  & 979  & 1279 & 1762 \\
\;\;\; \tilde t_1       &  &  300  & 300  & 300 & 506 \\
\;\;\; \tilde b_1       & &  690 & 648 & 716 & 1120 \\
\;\;\; \tilde \tau_1    &  &  607 & 433 & 316 & 295 \\
\hline \;\;\; a_\mu^{SUSY} \times 10^{10}  & 25.6 \; (16) & 2.95
& 2.91  & 2.75 & 1.04 \\
\hline \;\;\; \Omega_{\chi}h2  &  0.094 - 0.129
  & 0.095 & 0.095 & 0.115 & 0.101 \\
\;\;\; \sigsip (pb)\times 107  &   &  0.40  &  0.37  &  0.1 &  0.1 \\
\;\;\; {\rm BR} (B_s \rightarrow \mu^+\ \mu^- ) \times 107 & <
5 & 4.31  & 1.24  & 1.91 & 0.46 \\
\;\;\; {\rm BR} (B \rightarrow X_s \gamma)\times 104 & 3.41 \;(0.67)
& 7.49 & 10.01 & 4.09 & 0.62\\
\hline
\end{array}
$$
\end{table}


\section{Predictions and Summary} \label{sec:predictions}

This paper extends the analysis of a previous
paper~\cite{Dermisek:2003vn} to larger values of $m_{16}$, from 3 to 5
TeV, and larger CP odd Higgs mass $m_A \geq 500$ GeV. We find a
well--defined, narrow region of parameter space which provides the
observed relic density of dark matter, as well as a good fit to
precision electroweak data, including top, bottom and tau masses, and
acceptable bounds on the branching fraction of $B_s \rightarrow \mu^+\
\mu^-$.  We present predictions for Higgs and SUSY spectra (Table),
the dark matter detection cross section $\sigsip$ and the branching
ratio ${\rm BR}(B_s\rightarrow \mu^+\ \mu^-)$ in this region of
parameter space.

The MSO$_{10}$SM predicts relatively large first and second generation
scalar masses and smaller gaugino masses.
An immediate consequence of such heavy first and second generation sleptons
is the suppression of the SUSY contribution to the anomalous
magnetic moment of the muon. We find $a_\mu^{SUSY} \leq 3 \times 10^{-10}$ (see
Table).  This is consistent with the most recent experimental~\cite{gminus2}
and theoretical results at $1 \sigma$ if one uses
$\tau$--based analysis~\cite{Davier:2003pw}. However it is only consistent
with an $e^+e^-$--based analysis at 2 $\sigma$.

Another important consequence of the MSO$_{10}$SM is the large value
for $\tan\beta$ which leads to an enhanced branching ratio ${\rm BR}
(B_s \rightarrow \mu^+ \ \mu^-)$.  In addition, this is sensitive to
the value of the CP odd Higgs mass $m_{A}$~\cite{bsmumu}, scaling as
$m_{A}^{-4}$.  For $m_{A} = 500$ GeV, the branching ratio satisfies $2
\times 10^{-7} < {\rm BR} (B_s \rightarrow \mu^+ \ \mu^-) < 5 \times
10^{-7}$ for acceptable values of $\abundchi$ and $\chi^2 <
3$.\footnote{Where the upper limit is set by the recent DZero bound.}
In addition, we find that as $m_A$ increases, the region of parameter
space consistent with WMAP data is forced to larger values of
$M_{1/2}$ and smaller values of the light Higgs mass $m_h$.  Hence, we
find an upper bound on $m_A \approx 1.3$ TeV consistent with the light
Higgs mass bound $m_h > 114.4$ GeV.  For $m_A \leq 1.25$ TeV, we find
$BR(B_s \rightarrow \mu^+ \mu^-) > 10^{-8}$.  Hence all acceptable
regions of parameter space lead to observable rates for ${\rm BR}
(B_s\rightarrow \mu^+\ \mu^-)$.

We update our predictions for the cross section for elastic
neutralino--proton scattering due to scalar interactions $\sigsip$
as a function of the ${\rm BR} (B_s\rightarrow \mu^+\ \mu^-)$
(Fig.~\ref{fig:bsmm-sigp}) and the neutralino mass (Fig.~\ref{fig:sigp-mx})
for all regions satisfying the collider constraints,
$0.094 < \abundchi < 0.129$ and $\chi^2<3$.
For comparison, in Fig.~\ref{fig:sigp-mx}, we also show the bounds from the
present dark matter searches.
Over the next two to five years the experimental sensitivity is expected to
gradually improve by some three orders of magnitude.
This will cover large parts of the predicted ranges of
$\sigsip$.  Finally, we extend our previous analysis to include WIMP signals
in indirect detection.


\acknowledgments  We gratefully acknowledge the use of the GUT $\chi^2$
analysis code developed by T. Bla\v{z}ek.
R.D. is supported, in part, by the U.S. Department of Energy,
Contract DE-FG03-91ER-40674 and the Davis Institute for High Energy Physics.
R.Rda is supported by the program `Juan de la Cierva'
of the Ministerio de Educaci\'{o}n y Ciencia of Spain.
L.R. and R.RdA acknowledge support from ENTApP (European Network of Theoretical
Astroparticle Physics), member of ILIAS.
S.R. received partial support from DOE grant\# DOE/ER/01545-862.


\begin{thebibliography}{99}


\bibitem{cmssm}
G.~L.~Kane, C.~F.~Kolda, L.~Roszkowski and J.~D.~Wells,
Phys.\ Rev.\ D {\bf 49}, 6173 (1994) [arXiv:hep-ph/9312272].


\bibitem{Dermisek:2003vn}
R.~Dermisek, S.~Raby, L.~Roszkowski and R.~Ruiz De Austri,
JHEP {\bf 0304}, 037 (2003) [arXiv:hep-ph/0304101].


\bibitem{bdr}
T.~Blazek, R.~Dermisek and S.~Raby,
Phys.\ Rev.\ Lett.\  {\bf 88}, 111804 (2002) [arXiv:hep-ph/0107097];
Phys.\ Rev.\ D {\bf 65}, 115004 (2002) [arXiv:hep-ph/0201081].


\bibitem{wmap}
See, \eg, latest WMAP results in D.~N.~Spergel {\it et al.},
[arXiv:astro-ph/0302209].


\bibitem{bsmumu}
C.~Hamzaoui, M.~Pospelov and M.~Toharia,
Phys.\ Rev.\ D {\bf 59}, 095005 (1999) [arXiv:hep-ph/9807350];
K.~S.~Babu and C.~F.~Kolda,
Phys.\ Rev.\ Lett.\  {\bf 84}, 228 (2000) [arXiv:hep-ph/9909476];
P.~H.~Chankowski and L.~Slawianowska,
Phys.\ Rev.\ D {\bf 63}, 054012 (2001) [arXiv:hep-ph/0008046];
A.~Dedes, H.~K.~Dreiner and U.~Nierste,
Phys.\ Rev.\ Lett.\  {\bf 87}, 251804 (2001) [arXiv:hep-ph/0108037];
G.~Isidori and A.~Retico,
JHEP {\bf 0111}, 001 (2001) [arXiv:hep-ph/0110121];
A.~J.~Buras, P.~H.~Chankowski, J.~Rosiek and L.~Slawianowska,
Phys.\ Lett.\ B {\bf 546}, 96 (2002) [arXiv:hep-ph/0207241].


\bibitem{Bobeth:2001sq}
C.~Bobeth, T.~Ewerth, F.~Kruger and J.~Urban,
Phys.\ Rev.\ D {\bf 64}, 074014 (2001) [arXiv:hep-ph/0104284]; ibid.
Phys.\ Rev.\ D {\bf 66}, 074021 (2002) [arXiv:hep-ph/0204225].


\bibitem{Abazov:2004dj}
V.~M.~Abazov {\it et al.}  [D0 Collaboration],
arXiv:hep-ex/0410039.


\bibitem{cdfbsmm05prelim}
See CDF web page,
http://www-cdf.fnal.gov/physics/new/bottom/050407.blessed-bsmumu/.
For the previous published result, see D.~Acosta {\it et al.}  [CDF Collaboration],
Phys.\ Rev.\ Lett.\  {\bf 93}, 032001 (2004) [arXiv:hep-ex/0403032].


\bibitem{dzerobsmm05prelim}
See DZero web page, http://www-d0.fnal.gov/Run2Physics/WWW/results/prelim/B/B21/B21.pdf.


\bibitem{cdms04limit}
D.S.~Akerib, \etal, The CDMS Collaboration,
Phys.Rev.Lett. 93 (2004)
211301, [arXiv:astro-ph/0405033].


\bibitem{Tobe:2003bc}
K.~Tobe and J.~D.~Wells, [arXiv:hep-ph/0301015].


\bibitem{Auto:2003ys}
D.~Auto, H.~Baer, C.~Balazs, A.~Belyaev, J.~Ferrandis and X.~Tata,
[arXiv:hep-ph/0302155].


\bibitem{Balazs:2003mm}
C.~Balazs and R.~Dermisek,
arXiv:hep-ph/0303161.


\bibitem{Dermisek:2005ij}
R.~Dermisek and S.~Raby,
arXiv:hep-ph/0507045.


\bibitem{scrunching}  J.~A.~Bagger, J.~L.~Feng, N.~Polonsky and R.~J.~Zhang,
Phys.\ Lett.\ B {\bf 473}, 264 (2000) [arXiv:hep-ph/9911255].


\bibitem{masieroetal}  F.~Gabbiani, E.~Gabrielli, A.~Masiero and L.~Silvestrini,
Nucl.\ Phys.\ B {\bf 477}, 321 (1996) [arXiv:hep-ph/9604387];
T.~Besmer, C.~Greub, and T.~Hurth,
\npb{609}, 359 (2001) [arXiv:hep-ph/0105292].


\bibitem{or1+2}
K.~Okumura and L.~Roszkowski,
Phys.\ Rev.\ Lett.\  {\bf 92}, 161801 (2004) [arXiv:hep-ph/0208101];
JHEP {\bf 0310}, 024 (2003) [arXiv:hep-ph/0308102].


\bibitem{for1+2}
J.~Foster, K.~Okumura and L.~Roszkowski,
Phys.\ Lett.\ B {\bf 609}, 102 (2005)
[arXiv:hep-ph/0410323];
arXiv:hep-ph/0506146,
to appear in JHEP.


\bibitem{superk}  C.K. Jung talk, http://nngroup.physics.sunysb.edu/uno/UNO04-Keystone/.


\bibitem{pdecay} R.~Dermisek, A.~Mafi and S.~Raby,
Phys.\ Rev.\ D {\bf 63}, 035001 (2001) [arXiv:hep-ph/0007213].


\bibitem{chi2} T.~Blazek, M.~Carena, S.~Raby and C.~E.~Wagner,
Phys.\ Rev.\ D {\bf 56}, 6919 (1997) [arXiv:hep-ph/9611217].


\bibitem{carenaetal} H.~E.~Haber and R.~Hempfling,
Phys.\ Rev.\ D {\bf 48}, 4280 (1993) [arXiv:hep-ph/9307201];
 M.~Carena, J.~R.~Espinosa, M.~Quiros and C.~E.~Wagner,
Phys.\ Lett.\ B {\bf 355}, 209 (1995) [arXiv:hep-ph/9504316].


\bibitem{cqw} M.~Carena, M.~Quiros and C.~E.~Wagner,
Nucl.\ Phys.\ B {\bf 461}, 407 (1996) [arXiv:hep-ph/9508343].


\bibitem{pdg2000}  The Review of Particle Physics, D.~E.~Groom, et al.,
The European Physical Journal {\bf C15}, 1 (2000).


\bibitem{rhonew} P. Langacker, talk at Chicagoland seminar, October (1999).


\bibitem{Beneke:1999fe}
M.~Beneke and A.~Signer,
Phys.\ Lett.\ B {\bf 471}, 233 (1999) [arXiv:hep-ph/9906475].


\bibitem{nrr1+2}
T.~Nihei, L.~Roszkowski and R.~Ruiz de Austri,
JHEP {\bf 0105}, 063 (2001) [arXiv:hep-ph/0102308];
JHEP {\bf 0203}, 031 (2002) [arXiv:hep-ph/0202009].


\bibitem{eg97}
J.~Edsjo and P.~Gondolo,
Phys.\ Rev.\ D {\bf 56}, 1879 (1997) [arXiv:hep-ph/9704361].


\bibitem{nrr3}
T.~Nihei, L.~Roszkowski and R.~Ruiz de Austri,
JHEP {\bf 0207}, 024 (2002) [arXiv:hep-ph/0206266].


\bibitem{darksusy}
P.~Gondolo, J.~Edsjo, L.~Bergstrom, P.~Ullio, and T.~Baltz, \\
{\tt http://www.physto.se/edsjo/darksusy/}.


\bibitem{Arnowitt:2002cq}
R.~Arnowitt, B.~Dutta, T.~Kamon and M.~Tanaka,
Phys.\ Lett.\ B {\bf 538}, 121 (2002) [arXiv:hep-ph/0203069].


\bibitem{Ball:2000ba}
P.~Ball {\it et al.},
arXiv:hep-ph/0003238.


\bibitem{bkk05}
S.~Baek, Y.~G.~Kim and P.~Ko,
JHEP 0502 (2005) 067 [arXiv:hep-ph/0406033].


\bibitem{knrr1} Y.~G.~Kim, T.~Nihei, L.~Roszkowski and R.~Ruiz de Austri,
\jhep{0212}{034}{2002}, 
[arXiv:hep-ph/0208069] and work in progress.


\bibitem{jkg96}
G.~Jungman, M.~Kamionkowski and K.~Griest, \prep{267}{1996}{195}.


\bibitem{icecube}
J.~Ahrens, {\it et al.} [IceCube Collaboration],
Nucl. Phys. {\bf 118} ({\it Proc. Suppl.}) (2003) 388;
F.~Halzen, [arXiv:astro-ph/0311004];
F.~Halzen and D.~Hooper, JCAP{\bf 0401} (2004) 002.


\bibitem{isothermalmodel}
J.~Binney and S.~Tremaine, {\em Galactic Dynamics}
(Princeton University Press, Princeton, 1987.


\bibitem{nfwhalo95}
J.~F.~Navarro, C.~S.~Frenk and S.~D.~M. White,
\apj{462}{1996}{563} [arXiv:astro-ph/9508025] and
\apj{490}{1997}{493}.


\bibitem{morehalo99}
B.~Moore, S.~Ghigna, F.~Governato, G.~Lake, T.~Quinn, J.~Stadel and P.~Tozzi,
\apj{524}{1999}{19}.


\bibitem{bbmn04}
G.~Bertone, P.~Binetruy, Y.~Mambrini and E.~Nezri,
arXiv:hep-ph/0406083.

\bibitem{hooper+dingus05}
D.~Hooper and B.~Dingus,
Phys.\ Rev.\ D {\bf 70}, 113007 (2004) [arXiv:astro-ph/0210617].



\bibitem{hess}
F.~Aharonian {\it et al.} [HESS Collaboration],
Astron. Astrophys. {\bf 425}, L13 (2004) [arXiv:astro-ph/0408145].


\bibitem{beu99}
L.~Bergstrom, J.~Edsjo and P.~Ullio,
arXiv:astro-ph/9902012.


\bibitem{bess}
S.~Orito {\it et al.} [BESS Collaboration],
Phys.\ Rev.\ Lett.\ {\bf 84}, 1078 (2000).


\bibitem{fmw01}
J.L.~Feng, K.T.~Matchev and F.~Wilczek,
Phys.\ Rev.\ D {\bf 63}, 045024 (2001) [arXiv:astro-ph/0008115].


\bibitem{pamela}
M.~Pearce [Pamela Collaboration],
Nucl. Phys. {\bf 113} ({\it Proc. Suppl.}) (2002) 314.


\bibitem{ams}
J.~Casaus {\it et al.} [AMS Collaboration],
Nucl. Phys. {\bf 114} ({\it Proc. Suppl.}) (2003) 259.


\bibitem{gminus2} G.~W.~Bennett {\it et al.}  [Muon g-2 Collaboration],
Phys.\ Rev.\ Lett.\  {\bf 89}, 101804 (2002) [Erratum--ibid.\  {\bf
89}, 129903 (2002)] [arXiv:hep-ex/0208001].


\bibitem{Davier:2003pw}
M.~Davier, S.~Eidelman, A.~Hocker and Z.~Zhang,
Eur.\ Phys.\ J.\ C {\bf 31}, 503 (2003) [arXiv:hep-ph/0308213].



\end{thebibliography}
\end{document}